\documentclass[journal, 12pt, onecolumn, draftclsnofoot, letterpaper]{IEEEtran}
\usepackage{amsmath}
\usepackage{graphicx}
\usepackage{cite}
\usepackage{balance}
\usepackage{amssymb,amsfonts}

\usepackage{lipsum}
\usepackage{mathtools}
\usepackage{cuted}

\newtheorem{theorem}{\it Theorem}

\newtheorem{definition}{\it Definition}

\newtheorem{corollary}{\it Corollary}
\newtheorem{proposition}{\it Proposition}

\ifCLASSINFOpdf
\else
\fi
\begin{document}
%
\title{Channel Leakage, Information-Theoretic Limitations of Obfuscation, and Optimal Privacy Mask Design for Streaming Data
}
%
%

\author{Song~Fang~and~Quanyan~Zhu
\thanks{Song Fang and Quanyan Zhu are with the Department
of Electrical and Computer Engineering, New York University, New York, USA e-mails: song.fang@nyu.edu; quanyan.zhu@nyu.edu.}}

\maketitle

\begin{abstract}
	In this paper, we first introduce the notion of channel leakage as the minimum mutual information between the channel input and channel output. As its name indicates, channel leakage quantifies the minimum information leakage to the malicious receiver. In a broad sense, it can be viewed as a dual concept of channel capacity, which characterizes the maximum information transmission to the targeted receiver. We obtain explicit formulas of channel leakage for the white Gaussian case, the colored Gaussian case, and the fading case. We then utilize this notion to investigate the fundamental limitations of obfuscation in terms of privacy-distortion tradeoffs (as well as privacy-power tradeoffs) for streaming data; particularly, we derive analytical tradeoff equations for the stationary case, the non-stationary case, and the finite-time case. Our results also indicate explicitly how to design the privacy masks in an optimal way.
\end{abstract}


%
\IEEEpeerreviewmaketitle

%
%
%
%
\section{Introduction}

The topic of privacy for streaming data or dynamic data (see, e.g.,  \cite{venkitasubramaniam2015information, zhang2016dynamic, tanaka2017directed, han2018privacy, nekouei2019information, lu2019control, farokhi2020privacy, le2020differential} and the references therein) is attracting more and more attention in recent years. One important characteristic of streaming data or dynamic data is that information is contained not only in the samples at each time instant but also in the correlations over time \cite{Cov:06}. On the other hand, information-theoretic privacy (see, e.g., \cite{wyner1975wire, bloch2008wireless, liang2009information, liu2010securing, el2011network, bloch2011physical, sankar2011competitive, venkitasubramaniam2015information, han2016event, schaefer2017information, tanaka2017directed, nekouei2019information, farokhi2020privacy} and the references therein) features a fundamental privacy concept, and the most commonly used information-theoretic measure of privacy leakage is mutual information (see, e.g., \cite{wyner1975wire, bloch2008wireless, liang2009information, liu2010securing, el2011network, bloch2011physical, sankar2011competitive, venkitasubramaniam2015information, han2016event, schaefer2017information, tanaka2017directed, nekouei2019information, farokhi2020privacy} and the references therein). What we  discuss in this paper is in the general scope of information-theoretic privacy for streaming data (see, e.g., \cite{venkitasubramaniam2015information, tanaka2017directed, farokhi2020privacy} and the references therein). 


Particularly, in this paper we first formally introduce the concept of channel leakage, which is defined as the minimum mutual information between the channel input and channel output of a dynamic channel, supposing that the density function of the channel input is given while that of the channel noise can be designed, oftentimes subject to certain power constraints. For a given information source (as channel input) while subject to constraints on the information mask (as channel noise), channel leakage characterizes the minimum information leakage to a malicious receiver who has access to the masked version of the information source, i.e., the information source added with information mask (as channel output). We examine particularly the white Gaussian case, the colored Gaussian case, and the fading case, obtaining analytical formulas for the channel leakage as well as the optimal noise densities, indicating explicitly how to design the optimal information masks. It is also worth mentioning that, when designing the optimal information mask, channel leakage leads to ``obfuscating" power allocation policies, which are fundamentally different from the ``water-filling" policy for channel capacity as well as the ``reverse water-filling" policy for rate distortion.

Naturally, this notion of channel leakage may be employed to characterize the fundamental limitations of obfuscation in terms of privacy-distortion tradeoffs for streaming data. Specifically, consider the scenario in which a privacy mask is to be added to a given data stream, resulting in a masked data stream that an eavesdropper may have access to. The information-theoretic privacy leakage (on average) would then be defined as the mutual information rate between the original data stream (as a stochastic process) and the masked data stream (as another stochastic process), that is, how much information can be extracted from the latter about the former on average. Accordingly, we may ask two classes of questions. The first class is: Given a certain distortion constraint or power constraint, what would be the minimum  average privacy leakage in the long run, and how to design the privacy mask to achieve this lower bound? Or equivalently (in a dual manner): Given a certain requirement on the privacy level in terms of average privacy leakage, what would be the minimum distortion or the minimum power needed, and how to design the privacy mask to achieve this bound? It turns out the channel leakage and the ``obfuscating" power allocation provide mathematically explicit and physically intuitive solutions to all these problems, concerning stationary, non-stationary, as well as finite-time data streams.

Note that the analytical privacy-distortion tradeoffs obtained in this paper provide exact solutions to degraded versions of the privacy-utility tradeoff problems formulated in, e.g., \cite{rebollo2009t, du2012privacy, sankar2012smart, sankar2013utility, makhdoumi2013privacy}, for when the (Gaussian) useful information is a sufficient statistic of the private information. As a matter of fact, even when the useful information is not necessarily a sufficient statistic of the private information, the minimum information leakage rates (essentially from the useful information to the disclosed information) derived in this paper provide upper bounds on the minimum information leakage rates from the private information to the disclosed information considered in \cite{rebollo2009t, du2012privacy, sankar2012smart, sankar2013utility, makhdoumi2013privacy}. One may refer to Section~IV.A of \cite{makhdoumi2013privacy} for a more detailed discussion on this.


The rest of the paper is organized as follows. Section~II
introduces the technical preliminaries. In Section~III, we
introduce the notion of channel leakage and discuss its properties.
Section~IV presents the fundamental privacy-distortion tradeoffs and privacy-power tradeoffs. Conclusions are given in Section~V.




\section{Preliminaries}

Throughout the paper, we consider real-valued continuous random variables and random vectors, as well as discrete-time stochastic processes. All random variables, random vectors, and stochastic processes are assumed to be zero-mean. We represent random variables and random vectors using boldface letters. Given a stochastic process $\left\{ \mathbf{x}_{k}\right\}$, we denote the sequence $\mathbf{x}_0,\ldots,\mathbf{x}_{k}$ by $\mathbf{x}_{0,\ldots,k}$ for simplicity. The logarithm is with base $2$. A stochastic process $\left\{ \mathbf{x}_{k}\right\}, \mathbf{x}_k \in  \mathbb{R}$, is said to be  stationary if $ R_{\mathbf{x}}\left( i,k\right)=\mathbb{E}\left[  \mathbf{x}_i \mathbf{x}_{i+k} \right]$ depends only on $k$, and can thus be denoted as  $R_{\mathbf{x}}\left( k\right)$ for simplicity. The power spectrum of a stationary  process $\left\{ \mathbf{x}_{k} \right\}, \mathbf{x}_{k} \in \mathbb{R}$, is defined as
\begin{flalign}
S_{\mathbf{x}}\left( \omega\right)
=\sum_{k=-\infty}^{\infty} R_{\mathbf{x}}\left( k\right) \mathrm{e}^{-\mathrm{j}\omega k}. \nonumber
\end{flalign}
Moreover, the variance of $\left\{ \mathbf{x}_{k}\right\}$ is given by
\begin{flalign}
\sigma_{\mathbf{x}}^2
= \mathbb{E}\left[ \mathbf{x}_k^2 \right]
= \frac{1}{2\pi}\int_{-\pi}^{\pi} S_{\mathbf{x}}\left(\omega \right) \mathrm{d}  \omega. \nonumber
\end{flalign}


Entropy and mutual information are the most basic notions in information theory \cite{Cov:06}, which we introduce below.

\begin{definition} The differential entropy of a random vector $\mathbf{x} \in \mathbb{R}^m$ with density $p_{\mathbf{x}} \left(x\right)$ is defined as
	\begin{flalign}
	h\left( \mathbf{x} \right)
	=-\int p_{\mathbf{x}} \left(x\right) \log p_{\mathbf{x}} \left(x\right) \mathrm{d} x. \nonumber
	\end{flalign}
	The conditional differential entropy of random vector $\mathbf{x} \in \mathbb{R}^{m}$ given random vector $\mathbf{y} \in \mathbb{R}^{n}$ with joint density $p_{\mathbf{x}, \mathbf{y}} \left(x,y\right)$ and conditional density $p_{\mathbf{x} | \mathbf{y}} \left(x,y\right)$ is defined as
	\begin{flalign}
	h\left(\mathbf{x}\middle|\mathbf{y}\right)
	=-\int p_{\mathbf{x}, \mathbf{y}} \left(x,y\right)\log p_{\mathbf{x} | \mathbf{y}} \left(x,y\right) \mathrm{d}x\mathrm{d}y. \nonumber
	\end{flalign}
	The mutual information between random vectors $\mathbf{x} \in \mathbb{R}^{m}, \mathbf{y} \in \mathbb{R}^{n}$ with densities $p_{\mathbf{x}} \left(x\right)$, $p_{\mathbf{y}} \left( y \right) $ and joint density $p_{\mathbf{x}, \mathbf{y}} \left(x,y\right)$ is defined as
	\begin{flalign}
	I\left(\mathbf{x};\mathbf{y}\right)
	=\int p_{\mathbf{x}, \mathbf{y}} \left(x,y\right) \log \frac{p_{\mathbf{x}, \mathbf{y}} \left(x,y\right)}{p_{\mathbf{x}} \left(x\right) p_{\mathbf{y}} \left( y \right) }\mathrm{d}x\mathrm{d}y. \nonumber
	\end{flalign}
	The entropy rate of a stochastic process $\left\{ \mathbf{x}_{k}\right\},\mathbf{x}_{k}  \in \mathbb{R}^m$, is defined as
	\begin{flalign}
	h_\infty \left(\mathbf{x}\right)=\limsup_{k\to \infty} \frac{h\left(\mathbf{x}_{0,\ldots,k}\right)}{k+1}. \nonumber
	\end{flalign}
	The mutual information rate between two stochastic processes $\left\{ \mathbf{x}_{k}\right\},\mathbf{x}_{k}  \in \mathbb{R}^m$, and $\left\{ \mathbf{y}_{k}\right\},\mathbf{y}_{k}  \in \mathbb{R}^n$, is defined as
	\begin{flalign}
	I_{\infty}\left(\mathbf{x};\mathbf{y}\right)
	=\limsup_{k\to \infty} \frac{I \left(\mathbf{x}_{0,\ldots,k}; \mathbf{y}_{0,\ldots,k}\right)}{k+1}. \nonumber
	\end{flalign}
\end{definition}

Properties of these notions can be found in, e.g., \cite{Cov:06, yeung2008information, el2011network, fang2017towards}. 

On the other hand, consider an additive noise channel 
\begin{flalign} \label{channel}
\mathbf{y}_{k} = \mathbf{x}_{k} + \mathbf{z}_{k}, 
\end{flalign}
where $\left\{ \mathbf{x}_{k} \right\}, \mathbf{x}_{k} \in \mathbb{R}^m$, denotes the channel input, $\left\{ \mathbf{y}_{k} \right\}, \mathbf{y}_{k} \in \mathbb{R}^m$, denotes the channel output, and $\left\{ \mathbf{z}_{k} \right\}, \mathbf{z}_{k} \in \mathbb{R}^m$, denotes the additive noise. Suppose that $\left\{ \mathbf{z}_{k} \right\}$ is independent of $\left\{ \mathbf{x}_{k} \right\}$. For such a channel, the channel capacity is defined as follows \cite{Cov:06}.

\begin{definition}
	The channel capacity $C$ of the communication channel given in \eqref{channel} is defined as
	\begin{flalign}
	C
	= \sup_{p_{\mathbf{x}}} I_{\infty} \left( \mathbf{x}; \mathbf{y} \right)
	= \sup_{p_{\mathbf{x}}} \limsup_{k \to \infty} \frac{I \left( \mathbf{x}_{0, \ldots, k}; \mathbf{y}_{0, \ldots, k} \right)}{k+1}, \nonumber
	\end{flalign}
	where the supremum is taken over all possible densities $p_{\mathbf{x}}$ of the input process allowed for the channel.
\end{definition}


\section{Channel Leakage} \label{xyz}

Note first that channel leakage, like channel capacity, can be defined for classes of communication channels broader than additive noise channels. In this paper, however, we focus on additive channels for simplicity.

\begin{definition}
	The channel leakage $L$ of the communication channel given in \eqref{channel} is defined as
	\begin{flalign} \label{Ldef}
	L 
	= \inf_{p_{\mathbf{z}}} I_{\infty} \left( \mathbf{x}; \mathbf{y} \right)
	= \inf_{p_{\mathbf{z}}} \limsup_{k \to \infty} \frac{I \left( \mathbf{x}_{0, \ldots, k}; \mathbf{y}_{0, \ldots, k} \right)}{k+1},
	\end{flalign}
	where the infimum is taken over all possible densities $p_{\mathbf{z}}$ of the noise process allowed for the channel.
\end{definition}

%

As its name indicates, channel leakage quantifies the minimum information leakage to a malicious receiver.
In comparison, channel capacity characterizes the maximum information transmission to a targeted receiver.
In a broad sense, channel leakage may be viewed as a dual notion of channel capacity. Particularly, channel leakage is defined as the minimum mutual information rate between the channel input and channel output, with the channel input given; meanwhile, channel capacity is defined as the maximum mutual information rate between the channel input and channel output, with the channel noise given.

On the other hand, the following relationship between channel leakage and channel capacity may be established in general.

\begin{proposition}
	Denote the channel leakage with (given) input density $p_{\mathbf{x}}$ and noise power constraint $\mathbb{E} \left[ \mathbf{z}_{k}^2 \right] \leq N$ as
	\begin{flalign}
	L \left( p_{\mathbf{x}}, N \right)
	= \inf_{ \mathbb{E} \left[ \mathbf{z}_k^2 \right] \leq N} I_{\infty} \left( \mathbf{x} ; \mathbf{y} \right),
	\end{flalign} 
	and denote the channel capacity with (given) noise density $p_{\mathbf{z}}$ and input power constraint $\mathbb{E} \left[ \mathbf{x}_{k}^2 \right] \leq P$ as
	\begin{flalign}
	C \left( P, p_{\mathbf{z}} \right)
	= \sup_{ \mathbb{E} \left[ \mathbf{x}_k^2 \right] \leq P} I_{\infty} \left( \mathbf{x} ; \mathbf{y} \right).
	\end{flalign} 
	If
	\begin{flalign}
	\sigma_{\mathbf{x}}^2 
	= \int_{- \infty}^{\infty} x^2 p_{\mathbf{x}} \left( x \right) \mathrm{d} x
	= P,
	\end{flalign}
	and
	\begin{flalign}
	\sigma_{\mathbf{z}}^2 
	= \int_{- \infty}^{\infty} z^2 p_{\mathbf{z}} \left( z \right) \mathrm{d} z
	= N,
	\end{flalign}
	then
	\begin{flalign}
	L \left( p_{\mathbf{x}}, N \right) \leq C \left( P, p_{\mathbf{z}} \right).
	\end{flalign} 
\end{proposition}

\begin{IEEEproof}
	Since 
	\begin{flalign}
	\sigma_{\mathbf{x}}^2 
	= \int_{- \infty}^{\infty} x^2 p_{\mathbf{x}} \left( x \right) \mathrm{d} x
	= P, \nonumber
	\end{flalign}
	we have $\sigma_{\mathbf{x}}^2 \leq P$; similarly, since
	\begin{flalign}
	\sigma_{\mathbf{z}}^2 
	= \int_{- \infty}^{\infty} z^2 p_{\mathbf{z}} \left( z \right) \mathrm{d} z
	= N, \nonumber
	\end{flalign}
	we have $\sigma_{\mathbf{z}}^2 \leq N$. As a result,
	\begin{flalign}
	L \left( p_{\mathbf{x}}, N \right)
	= \inf_{ \mathbb{E} \left[ \mathbf{z}_k^2 \right] \leq N} I_{\infty} \left( \mathbf{x} ; \mathbf{y} \right)
	\leq  I_{\infty} \left( \mathbf{x} ; \mathbf{y} \right) |_{p_{\mathbf{x}},p_{\mathbf{z}}}\leq \sup_{ \mathbb{E} \left[ \mathbf{x}_k^2 \right] \leq P} I_{\infty} \left( \mathbf{x} ; \mathbf{y} \right)
	= C \left( P, p_{\mathbf{z}} \right),
	\nonumber
	\end{flalign} 
	which concludes the proof.
\end{IEEEproof}

Let us now consider some special classes of communication channels. We shall start with the white Gaussian case.

\begin{theorem} \label{AWGI}
	Consider a scalar channel of $m=1$ and suppose that the channel input $\left\{ \mathbf{x}_{k} \right\}$ is stationary white Gaussian with variance $\sigma_{\mathbf{x}}^2$. Suppose also that $\left\{ \mathbf{z}_{k} \right\}$ is independent of $\left\{ \mathbf{x}_{k} \right\}$. Then, the channel leakage with noise power constraint $\mathbb{E} \left[ \mathbf{z}_k^2 \right] \leq N$ is given by
	\begin{flalign}
	L
	= \frac{1}{2} \log \left( 1 + \frac{\sigma_{\mathbf{x}}^2}{N} \right).
	\end{flalign}
\end{theorem}

\begin{IEEEproof}
	See Appendix~A.
\end{IEEEproof}


It is known \cite{Cov:06} that the channel capacity of a scalar additive white Gaussian noise channel, where the channel noise $\left\{ \mathbf{z}_{k} \right\}$ is stationary white Gaussian with variance $\sigma_{\mathbf{z}}^2$ and the input power constraint is $\mathbb{E} \left[ \mathbf{x}_k^2 \right] \leq P$, is given by
\begin{flalign}
C
= \frac{1}{2} \log \left( 1 + \frac{P}{\sigma_{\mathbf{z}}^2} \right). \nonumber
\end{flalign} 

Note that the distribution of a zero-mean stationary white Gaussian process is fully determined by its variance (second moment) \cite{Pap:02}.
As such, if $\sigma_{\mathbf{x}}^2 = P$ and $\sigma_{\mathbf{z}}^2 = N$, then it holds for this pair that
\begin{flalign}
L \left( p_{\mathbf{x}}, N \right) 
= L \left( \sigma_{\mathbf{x}}^2, N \right) 
= \frac{1}{2} \log \left( 1 + \frac{P}{N} \right) 
= C \left( P, \sigma_{\mathbf{z}}^2 \right) 
= C \left( P, p_{\mathbf{z}} \right).
\end{flalign} 


Let us next consider the colored Gaussian case and present the following theorem.

\begin{theorem} \label{ACGI}
	Consider a scalar channel of $m=1$ and suppose that the channel input $\left\{ \mathbf{x}_{k} \right\}$ is stationary colored Gaussian with power spectrum $S_{\mathbf{x}} \left( \omega \right)$. Suppose also that $\left\{ \mathbf{z}_{k} \right\}$ is independent of $\left\{ \mathbf{x}_{k} \right\}$. Then, the channel leakage with noise power constraint $\mathbb{E} \left[ \mathbf{z}_k^2 \right] \leq N$ is given by
	\begin{flalign}
	L 
	= \frac{1}{2 \pi} \int_{0}^{2 \pi} \log \sqrt{ 1 + \frac{S_{\mathbf{x}} \left( \omega \right)}{N \left( \omega \right)} } \mathrm{d} \omega,
	\end{flalign} 
	where 
	\begin{flalign} \label{fire}
	N \left( \omega \right) = \frac{\zeta}{2 \left[ 1 + \sqrt{1 + \frac{\zeta}{S_{\mathbf{x}} \left( \omega \right)}} \right]},
	\end{flalign} 
	and $\zeta \geq 0$ satisfies
	\begin{flalign} \label{fire2}
	\frac{1}{2 \pi} \int_{0}^{2 \pi} N \left( \omega \right) \mathrm{d} \omega 
	= \frac{1}{2 \pi} \int_{0}^{2 \pi} \frac{\zeta}{2 \left[ 1 + \sqrt{1 + \frac{\zeta}{S_{\mathbf{x}} \left( \omega \right)}} \right]} \mathrm{d} \omega 
	= N.
	\end{flalign} 
\end{theorem}

\begin{IEEEproof}
	See Appendix~B.
\end{IEEEproof}

The power allocation policy in \eqref{fire} and \eqref{fire2} may be viewed as an ``obfuscating" power allocation policy (in the spectral/frequency domain), referring to a policy that delivers more power to frequencies that are with larger (input) spectra, which is opposite to the ``water-filling" policy for channel capacity \cite{Cov:06}. 
Particularly, the channel capacity of a scalar additive colored Gaussian noise channel, where the channel noise $\left\{ \mathbf{z}_{k} \right\}$ is stationary colored Gaussian with power spectrum $S_{\mathbf{z}} \left( \omega \right)$ and the input power constraint is $\mathbb{E} \left[ \mathbf{x}_k^2 \right] \leq P$, is given by \cite{Cov:06}
\begin{flalign}
C
= \frac{1}{2 \pi} \int_{0}^{2 \pi} \log \sqrt{ 1 + \frac{P \left( \omega \right)}{S_{\mathbf{z}} \left( \omega \right)} } \mathrm{d} \omega, \nonumber
\end{flalign} 
where 
\begin{flalign} \label{water}
P \left( \omega \right) 
= \max \left\{ 0, \zeta - S_{\mathbf{z}} \left( \omega \right) \right\}, 
\end{flalign} 
and $\zeta \geq 0$ satisfies
\begin{flalign} \label{water2}
\frac{1}{2 \pi} \int_{0}^{2 \pi} P \left( \omega \right) \mathrm{d} \omega 
= \frac{1}{2 \pi} \int_{0}^{2 \pi} \max \left\{ 0, \zeta - S_{\mathbf{z}} \left( \omega \right) \right\} \mathrm{d} \omega 
= P.
\end{flalign} 
Note that the power allocation policy given in \eqref{water} and \eqref{water2} is also known as ``water-filling" (in the spectral/frequency domain) \cite{Cov:06}.

On the other hand, the distribution of a zero-mean stationary colored Gaussian process is fully determined by its power spectrum (essentially second moments) \cite{Pap:02}.
As such, if
\begin{flalign}
\sigma_{\mathbf{x}}^2 
= \frac{1}{2 \pi} \int_{0}^{2 \pi} S_{\mathbf{x}} \left( \omega \right) \mathrm{d} \omega 
= P,
\end{flalign}
and
\begin{flalign}
\sigma_{\mathbf{z}}^2 
= \frac{1}{2 \pi} \int_{0}^{2 \pi} S_{\mathbf{z}} \left( \omega \right) \mathrm{d} \omega 
= N,
\end{flalign}
then it holds for this pair that
\begin{flalign}
L \left( p_{\mathbf{x}}, N \right) 
&= L \left( S_{\mathbf{x}} \left( \omega \right), N \right) 
= \frac{1}{2 \pi} \int_{0}^{2 \pi} \log \sqrt{ 1 + \frac{S_{\mathbf{x}} \left( \omega \right)}{N \left( \omega \right)} } \mathrm{d} \omega 
\leq \frac{1}{2 \pi} \int_{0}^{2 \pi} \log \sqrt{ 1 + \frac{S_{\mathbf{x}} \left( \omega \right)}{S_{\mathbf{z}} \left( \omega \right)} } \mathrm{d} \omega \nonumber \\
&
\leq \frac{1}{2 \pi} \int_{0}^{2 \pi} \log \sqrt{ 1 + \frac{P \left( \omega \right)}{S_{\mathbf{z}} \left( \omega \right)} } \mathrm{d} \omega 
= C \left( P, S_{\mathbf{z}} \left( \omega \right) \right) 
= C \left( P, p_{\mathbf{z}} \right),
\end{flalign} 
where $N \left( \omega \right)$ and $P \left( \omega \right)$ are given by \eqref{fire} and \eqref{water}, respectively. In fact, it may be further verified that
\begin{flalign}
L \left( S_{\mathbf{x}} \left( \omega \right), N \right)
= C \left( P, S_{\mathbf{z}} \left( \omega \right) \right),
\end{flalign} 
if and only if $S_{\mathbf{x}} \left( \omega \right)$ (in $L \left( S_{\mathbf{x}} \left( \omega \right), N \right)$) and $S_{\mathbf{z}} \left( \omega \right)$ (in $C \left( P, S_{\mathbf{z}} \left( \omega \right) \right)$) are constants; that is, $\left\{ \mathbf{x}_{k} \right\}$ is white (in the definition of channel leakage) while $\left\{ \mathbf{z}_{k} \right\}$ is white (in the definition of channel capacity).

The next corollary follows as a special case of Theorem~\ref{ACGI} (particularly, see its proof in Appendix~B).

\begin{corollary} \label{ACGIc}
	Consider $m$ parallel (independent) channels with
	\begin{flalign}
	\mathbf{y} = \mathbf{x} + \mathbf{z},
	\end{flalign}
	where $ \mathbf{x},\mathbf{y},\mathbf{z} \in  \mathbb{R}^m$, and $ \mathbf{z} $ is independent of $ \mathbf{x} $. In addition, $\mathbf{x}$ is Gaussian with covariance \begin{flalign}
	\Sigma_{\mathbf{x}} = \mathrm{diag} \left( \sigma_{\mathbf{x} \left( 1 \right)}^2, \ldots, \sigma_{\mathbf{x} \left( m \right)}^2 \right),
	\end{flalign} 
	where $\mathbf{x} \left( i \right), i = 1, \ldots, m$, denotes the $i$-th element of $\mathbf{x}$, and $\sigma_{\mathbf{x} \left( i \right)}^2$ denotes its variance.
	Suppose that the noise power constraint is given by 
	\begin{flalign} 
	\mathrm{tr} \left(\Sigma_{\mathbf{z} }  \right)
	= \mathbb{E} \left[ \sum_{i=1}^{m}
	\mathbf{z}^{2}\left( i \right) \right] \leq N, 
	\end{flalign}
	where $\mathbf{z} \left( i \right)$ denotes the $i$-th element of $\mathbf{z}$. Then, the channel leakage is given by
	\begin{flalign}
	L 
	= \sum_{i=1}^{m} \frac{1}{2}\log \left[ 1+\frac{\sigma_{\mathbf{x} \left( i \right)}^2}{N_{i}} \right],
	\end{flalign}
	where 
	\begin{flalign} 
	N_{i}=\frac{\zeta}{2\left[ \sqrt{1+\frac{\zeta }{\sigma_{\mathbf{x} \left( i \right)}^2}}+1 \right]},
	\end{flalign}
	with $\zeta \geq 0$ satisfying
	\begin{flalign} 
	\sum_{i=1}^{m}  N_{i}
	= 	\sum_{i=1}^{m} \frac{\zeta}{2\left[ \sqrt{1+\frac{\zeta }{\sigma_{\mathbf{x} \left( i \right)}^2}}+1 \right]} = N.
	\end{flalign}
\end{corollary}

The aim that we single out this result is to present a side-by-side comparison between the ``obfuscating" policy for channel leakage and the ``reverse water-filling" policy for rate distortion \cite{Cov:06} (see the subsequent paragraph for details); on the other hand, we already showed the difference between the ``obfuscating" policy for channel leakage and the ``water-filling" policy for channel capacity in the discussions after Theorem~\ref{ACGI}. As such, channel leakage is seen to be essentially different from rate distortion as well (in addition to channel capacity).

Particularly, consider a parallel Gaussian source with $m$ independent Gaussian random variables $\mathbf{x}_1,\ldots,\mathbf{x}_m$. Suppose that the variances of $\mathbf{x}_1,\ldots,\mathbf{x}_m$ are  $\sigma_{1}^2, \ldots, \sigma_{m}^2$, respectively, and the distortion measure is
$\sum_{i=1}^{m} \left(\widehat{\mathbf{x}}_{i}-\mathbf{x}_{i}\right)^{2}$.
Then, the rate distortion function is
\begin{flalign} 
R\left( D \right) 
= \sum_{i=1}^{m} \frac{1}{2} \log \frac{\sigma_{i}^2}{D_{i}},\nonumber
\end{flalign}
where the distortion $D_{i}$ for  $\mathbf{x}_i$ is given by
\begin{flalign}
D_{i} = \begin{cases}
\zeta, & \text{if $\zeta < \sigma_{i}^2,$}\\
\sigma_{i}^2, & \text{if $\zeta \geq \sigma_{i}^2,$}
\end{cases} \nonumber
\end{flalign}
and $\zeta$ satisfies
\begin{flalign}
\sum_{i=1}^{m} D_{i} =D.  \nonumber
\end{flalign}
On the other hand, the variance of $\widehat{\mathbf{x}}_i$ is given by
\begin{flalign} 
\widehat{\sigma}_{i}^2 = \sigma_{i}^2 - D_{i} = \begin{cases}
\sigma_{i}^2 - \zeta, & \text{if $\zeta < \sigma_{i}^2,$}\\
0, & \text{if $\zeta \geq \sigma_{i}^2.$}
\end{cases} \nonumber
\end{flalign}
This allocation policy is also known as the ``reverse water-filling" \cite{Cov:06}. (See also the remarks later after Theorem~\ref{finitetime} for additional discussions on the comparison.)

In parallel, the power constraint might be imposed on the channel output. In this case, we first present the following result for the colored Gaussian case.

\begin{theorem} \label{output1}
	Consider a scalar channel of $m=1$ and suppose that the channel input $\left\{ \mathbf{x}_{k} \right\}$ is stationary colored Gaussian with power spectrum $S_{\mathbf{x}} \left( \omega \right)$. Suppose also that $\left\{ \mathbf{z}_{k} \right\}$ is independent of $\left\{ \mathbf{x}_{k} \right\}$. Then, the channel leakage with output power constraint $\mathbb{E} \left[ \mathbf{y}_k^2 \right] \leq Y$ is given by
	\begin{flalign}
	L 
	= \frac{1}{2 \pi} \int_{0}^{2 \pi} \log \sqrt{ 1 + \frac{S_{\mathbf{x}} \left( \omega \right)}{N \left( \omega \right)} } \mathrm{d} \omega,
	\end{flalign} 
	where 
	\begin{flalign} 
	N \left( \omega \right) = \frac{\zeta}{2 \left[ 1 + \sqrt{1 + \frac{\zeta}{S_{\mathbf{x}} \left( \omega \right)}} \right]},
	\end{flalign} 
	and $\zeta \geq 0$ satisfies
	\begin{flalign}
	\frac{1}{2 \pi} \int_{0}^{2 \pi} N \left( \omega \right) \mathrm{d} \omega 
	= \frac{1}{2 \pi} \int_{0}^{2 \pi} \frac{\zeta}{2 \left[ 1 + \sqrt{1 + \frac{\zeta}{S_{\mathbf{x}} \left( \omega \right)}} \right]} \mathrm{d} \omega 
	= Y - \frac{1}{2 \pi} \int_{0}^{2 \pi} S_{\mathbf{x}} \left( \omega \right) \mathrm{d} \omega.
	\end{flalign} 
\end{theorem}

Note that Theorem~\ref{output1} is essentially equivalent to the channel leakage with noise power constraint 
\begin{flalign}
\mathbb{E} \left[ \mathbf{z}_k^2 \right]
\leq Y - \frac{1}{2 \pi} \int_{0}^{2 \pi} S_{\mathbf{x}} \left( \omega \right) \mathrm{d} \omega.
\end{flalign} 
Specifically, since $\left\{ \mathbf{z}_{k} \right\}$ is independent of $\left\{ \mathbf{x}_{k} \right\}$, we have $S_{\mathbf{y}} \left( \omega \right) = S_{\mathbf{x} + \mathbf{z}} \left( \omega \right) = S_{\mathbf{x}} \left( \omega \right) + N \left( \omega \right)$, and thus
\begin{flalign}
\mathbb{E} \left[ \mathbf{z}_k^2 \right]
&= \frac{1}{2 \pi} \int_{0}^{2 \pi} N \left( \omega \right) \mathrm{d} \omega 
= \frac{1}{2 \pi} \int_{0}^{2 \pi} S_{\mathbf{y}} \left( \omega \right) \mathrm{d} \omega  - \frac{1}{2 \pi} \int_{0}^{2 \pi} S_{\mathbf{x}} \left( \omega \right) \mathrm{d} \omega \nonumber \\
&= \mathbb{E} \left[ \mathbf{y}_k^2 \right] - \frac{1}{2 \pi} \int_{0}^{2 \pi} S_{\mathbf{x}} \left( \omega \right) \mathrm{d} \omega 
\leq Y - \frac{1}{2 \pi} \int_{0}^{2 \pi} S_{\mathbf{x}} \left( \omega \right) \mathrm{d} \omega.
\end{flalign} 

In both the white Gaussian case and the colored Gaussian case, the processes are stationary. Let us now consider the non-stationary fading model (as inspired by the fading channel \cite{goldsmith2005wireless, Tse:04}, but with subtle differences). Particularly, consider a scalar channel of $m=1$ with   
\begin{flalign} \label{fadingchannel}
\mathbf{y}_k
= \mathbf{x}_k + \mathbf{z}_k
= \mathbf{h}_k \widehat{\mathbf{x}}_k + \mathbf{z}_k,
\end{flalign}
where $\left\{ \widehat{\mathbf{x}}_{k} \right\}, \widehat{\mathbf{x}}_k \in \mathbb{R}$, is assumed to be stationary white Gaussian with variance $\sigma_{\widehat{\mathbf{x}}}^2$,  $\left\{ \mathbf{h}_k \right\}, \mathbf{h}_k \in \mathbb{R}$, is the fading process, and $\left\{ \mathbf{z}_{k} \right\}, \mathbf{z}_k \in \mathbb{R}$, is  the noise. Note that $\left\{ \mathbf{x}_{k} \right\}$ is no longer stationary, since the  variance $\sigma_{\mathbf{x}_{k}}^2 = \left| h_{k} \right|^2 \sigma_{\widehat{\mathbf{x}}}^2 $ of $\mathbf{x}_{k}$ is in general time varying, as a function of $ h_{k}$ which denotes the realization of $ \mathbf{h}_{k}$.

Assume further that $\left\{ \widehat{\mathbf{x}}_{k} \right\}$, $\left\{ \mathbf{h}_{k} \right\}$, and $\left\{ \mathbf{z}_{k} \right\}$ are mutually independent. Suppose that the fading time of $\left\{ \mathbf{h}_k \right\}$ is relatively small compared to transmission duration, i.e., the codeword length spans many coherence periods \cite{goldsmith2005wireless, Tse:04}. We say that this fading model is with noise side information if the noise designer has access to the realizations of $\left\{ \mathbf{h}_k \right\}$, denoted as $\left\{ h_k \right\}$, at each time $k$ (cf. the transmitter side information in fading channels \cite{goldsmith2005wireless, Tse:04}).

\begin{theorem} \label{fading}
	Consider the fast fading channel given in \eqref{fadingchannel}. We may consider the cases with and without noise side information separately.
	
	{\em Case~1: Without noise side information.} The channel leakage with noise power constraint $\mathbb{E} \left[ \mathbf{z}_{k}^2 \right] \leq N$ is given by
	\begin{flalign}
	L=\mathbb{E} \left[\frac{1}{2} \log \left( 1+\frac{\left| \mathbf{h}_k \right|^{2} \sigma_{\mathbf{x}}^2 }{N} \right) \right].
	\end{flalign}
	
	{\em Case~2: With noise side information.} The channel leakage with noise power constraint $\mathbb{E} \left[ \mathbf{z}_{k}^2 \right] \leq N$ is given by
	\begin{flalign}
	L = \mathbb{E} \left[ \frac{1}{2}\log
	\left(1+\frac{ \left| \mathbf{h}_k \right|^2 \sigma_{\mathbf{x}}^2}{N_{k}}\right) \right],
	\end{flalign}
	where
	\begin{flalign}
	N_{k} = \frac{\zeta}{2\left(\sqrt{1+\frac{\zeta }{\left| h_k \right|^2 \sigma_{\mathbf{x}}^2}}+1
		\right)}, 
	\end{flalign}
	and $\zeta \geq 0$ satisfies
	\begin{flalign}
	\mathbb{E} \left[   N_{k} \right]
	= \mathbb{E} \left[   \frac{\zeta}{2\left(\sqrt{1+\frac{\zeta }{\left| \mathbf{h}_k \right|^2 \sigma_{\mathbf{x}}^2}}+1
		\right)} \right]
	=N.
	\end{flalign}
\end{theorem}

\begin{IEEEproof}
	See Appendix~C.
\end{IEEEproof}


To compare with channel capacity, consider a fast fading channel \cite{goldsmith2005wireless, Tse:04} where the noise $\left\{ \mathbf{z}_k \right\}$ is stationary white Gaussian with variance $\sigma_{\mathbf{z}}^2$. If the fading channel is without transmitter side information, then the channel capacity with input power constraint $\mathbb{E} \left[ \widehat{\mathbf{x}}_{k}^2 \right] \leq P$ is given by
\begin{flalign}
C= \mathbb{E} \left[\frac{1}{2} \log \left( 1+\frac{\left|\mathbf{h}_k\right|^{2} P}{\sigma_{\mathbf{z}}^2} \right) \right]. \nonumber
\end{flalign}
If the fading channel is with receiver side information, then the channel capacity with input power constraint $\mathbb{E} \left[ \widehat{\mathbf{x}}_{k}^2 \right] \leq P$ is given by 
\begin{flalign}
C=\mathbb{E} \left[\frac{1}{2} \log \left( 1+\frac{ \left|\mathbf{h}_k\right|^{2}P_k}{\sigma_{\mathbf{z}}^2} \right)\right],\nonumber
\end{flalign}
where
\begin{flalign}
P_k =\max \left\{0,\zeta- \frac{\sigma_{\mathbf{z}}^2}{\left| h_k\right|^{2}}  \right\},\nonumber&
\end{flalign}
and $\zeta \geq 0$ satisfies
\begin{flalign}
\mathbb{E} \left[ P_k \right] =P.\nonumber
\end{flalign}

When the power constraint is imposed on the channel output, we can obtain the following result.

\begin{theorem} \label{output2}
	Consider the fast fading channel given in \eqref{fadingchannel}. 
	
	{\em Case~1: Without noise side information.} The channel leakage with output power constraint $\mathbb{E} \left[ \mathbf{y}_k^2 \right] \leq Y$ is given by
	\begin{flalign}
	L=\mathbb{E} \left[\frac{1}{2} \log \left( 1+\frac{\left| \mathbf{h}_k \right|^{2} \sigma_{\mathbf{x}}^2 }{Y - \mathbb{E} \left[ \left| \mathbf{h}_k \right|^2 \right] \sigma_{\mathbf{x}}^2} \right) \right].
	\end{flalign}
	
	{\em Case~2: With noise side information.} The channel leakage with output power constraint $\mathbb{E} \left[ \mathbf{y}_k^2 \right] \leq Y$ is given by
	\begin{flalign}
	L = \mathbb{E} \left[ \frac{1}{2}\log
	\left(1+\frac{ \left| \mathbf{h}_k \right|^2 \sigma_{\mathbf{x}}^2}{N_{k}}\right) \right],
	\end{flalign}
	where
	\begin{flalign}
	N_{k} = \frac{\zeta}{2\left(\sqrt{1+\frac{\zeta }{\left| h_k \right|^2 \sigma_{\mathbf{x}}^2}}+1
		\right)},
	\end{flalign}
	and $\zeta \geq 0$ satisfies
	\begin{flalign}
	\mathbb{E} \left[   N_{k} \right]
	= \mathbb{E} \left[   \frac{\zeta}{2\left(\sqrt{1+\frac{\zeta }{\left| \mathbf{h}_k \right|^2 \sigma_{\mathbf{x}}^2}}+1
		\right)} \right]
	= Y - \mathbb{E} \left[ \left| \mathbf{h}_k \right|^2 \right] \sigma_{\mathbf{x}}^2.
	\end{flalign}
\end{theorem}

Note that Theorem~\ref{output2} is essentially equivalent to the channel leakage with noise power constraint 
\begin{flalign}
\mathbb{E} \left[ \mathbf{z}_k^2 \right]
\leq Y - \mathbb{E} \left[ \left| \mathbf{h}_k \right|^2 \right] \sigma_{\mathbf{x}}^2.
\end{flalign} 
Particularly, since $\left\{ \mathbf{z}_{k} \right\}$, $\left\{ \mathbf{x}_{k} \right\}$, and $\left\{ \mathbf{h}_{k} \right\}$ are mutually independent, we have 
\begin{flalign}
\mathbb{E} \left[ \mathbf{y}_k^2 \right] 
&= \mathbb{E} \left[  \left( \mathbf{h}_k \mathbf{x}_k + \mathbf{z}_k \right)^2 \right] 
= \mathbb{E} \left[  \left( \mathbf{h}_k \mathbf{x}_k \right)^2 \right] + \mathbb{E} \left[ \mathbf{z}_k^2 \right] 
= \mathbb{E} \left[  \left| \mathbf{h}_k \right|^2 \mathbf{x}_k^2 \right] + \mathbb{E} \left[ \mathbf{z}_k^2 \right] \nonumber \\
&= \mathbb{E} \left[  \left| \mathbf{h}_k \right|^2 \right] \mathbb{E} \left[ \mathbf{x}_k^2 \right] + \mathbb{E} \left[ \mathbf{z}_k^2 \right] 
= \mathbb{E} \left[  \left| \mathbf{h}_k \right|^2 \right] \sigma_{\mathbf{x}}^2 + \mathbb{E} \left[ \mathbf{z}_k^2 \right]
,
\end{flalign} 
and thus
\begin{flalign}
\mathbb{E} \left[ \mathbf{z}_k^2 \right]
= \mathbb{E} \left[ \mathbf{y}_k^2 \right] - \mathbb{E} \left[  \left| \mathbf{h}_k \right|^2 \right] \sigma_{\mathbf{x}}^2
\leq Y - \mathbb{E} \left[ \left| \mathbf{h}_k \right|^2 \right] \sigma_{\mathbf{x}}^2.
\end{flalign}

	\section{Fundamental Limitations of Obfuscation and Optimal Privacy Mask Design}

In this section, we present the fundamental lower bounds on the information leakage rate of streaming data. Specifically, consider the scenario in which a privacy mask is to be added to a given data stream, resulting in a masked data stream that an eavesdropper may have access to. The information-theoretic privacy leakage (on average) would then be defined as the mutual information rate between the original data stream (as a stochastic process) and the masked data stream (as another stochastic process), that is, how much information can be extracted from the latter about the former on average. Accordingly, we may ask two classes of questions. The first class is: Given a certain distortion constraint or power constraint, what would be the minimum average privacy leakage in the long run, and how to design the privacy mask to achieve this lower bound? Or equivalently (in a dual manner): Given a certain requirement on the privacy level in terms of average privacy leakage, what would be the minimum distortion or the minimum power needed, and how to design the privacy mask to achieve this bound? We shall address these questions one by one.

\subsection{Stationary Data Stream}

In this subsection, we consider stationary data streams.
We first consider the case when there exists a constraint on the distortion between the original data and the masked data.

\begin{theorem} \label{privacy1}
	Consider a data stream $\left\{ \mathbf{x}_{k} \right\}, \mathbf{x}_{k} \in \mathbb{R}$. Suppose that $\left\{ \mathbf{x}_{k} \right\}$ is stationary colored Gaussian with power spectrum $S_{\mathbf{x}} \left( \omega \right)$. For the sake of privacy, a noise $\left\{ \mathbf{n}_{k} \right\}, \mathbf{n}_{k} \in \mathbb{R}$, that is independent of $\left\{ \mathbf{x}_{k} \right\}$ is to be added to $\left\{ \mathbf{x}_{k} \right\}$ as its privacy mask, resulting in a masked streaming data $\left\{ \overline{\mathbf{x}}_{k} \right\}, \overline{\mathbf{x}}_{k} = \mathbf{x}_{k} + \mathbf{n}_{k}$,  whereas the properties of $\left\{ \mathbf{n}_{k} \right\}$ can be designed subject to a distortion constraint 
	\begin{flalign}
	\mathbb{E} \left[ \left( \mathbf{x}_k - \overline{\mathbf{x}}_{k} \right)^2 \right] \leq D.
	\end{flalign} 
	Then, in order to minimize the  information leakage rate
	\begin{flalign}
	I_{\infty} \left( \mathbf{x} ; \overline{\mathbf{x}} \right),
	\end{flalign}
	the noise
	$\left\{ \mathbf{n}_{k} \right\}$ should be chosen as a stationary colored Gaussian process. In addition, the power spectrum of $\left\{ \mathbf{n}_{k} \right\}$ should be chosen as
	\begin{flalign} \label{spectrum1}
	N \left( \omega \right) = \frac{\zeta}{2 \left[ 1 + \sqrt{1 + \frac{\zeta}{S_{\mathbf{x}} \left( \omega \right)}} \right]},
	\end{flalign} 
	where $\zeta \geq 0$ satisfies
	\begin{flalign}
	\frac{1}{2 \pi} \int_{0}^{2 \pi} N \left( \omega \right) \mathrm{d} \omega 
	= \frac{1}{2 \pi} \int_{0}^{2 \pi} \frac{\zeta}{2 \left[ 1 + \sqrt{1 + \frac{\zeta}{S_{\mathbf{x}} \left( \omega \right)}} \right]} \mathrm{d} \omega 
	= D.
	\end{flalign} 
	Correspondingly, the minimum information leakage rate is given by
	\begin{flalign} \label{leakage1}
	\inf_{\mathbb{E} \left[ \left( \mathbf{x}_k - \overline{\mathbf{x}}_{k} \right)^2 \right] \leq D} I_{\infty} \left( \mathbf{x} ; \overline{\mathbf{x}} \right)
	= L \left( S_{\mathbf{x}} \left( \omega \right), D \right) 
	= \frac{1}{2 \pi} \int_{0}^{2 \pi} \log \sqrt{ 1 + \frac{S_{\mathbf{x}} \left( \omega \right)}{N \left( \omega \right)} } \mathrm{d} \omega.
	\end{flalign} 
\end{theorem}

\begin{IEEEproof}
	Note first that the distortion constraint $
	\mathbb{E} \left[ \left( \mathbf{x}_k - \overline{\mathbf{x}}_{k} \right)^2 \right] \leq D $
	is equivalent to being with a power constraint $\mathbb{E} \left[ \mathbf{n}_k^2 \right] \leq D$, since $\overline{\mathbf{x}}_{k} = \mathbf{x}_{k} + \mathbf{n}_k $ and thus $\left( \mathbf{x}_k - \overline{\mathbf{x}}_{k} \right)^2 = \mathbf{n}_k^2$.
	The rest of the proof proceeds as in the that of Theorem~\ref{ACGI}, by viewing $\mathbf{n}_{k}$ and $\overline{\mathbf{x}}_{k}$ as $\mathbf{z}_{k}$ and $\mathbf{y}_{k}$ therein, respectively. 
	That is to say, $\left\{ \mathbf{n}_{k} \right\}$ should be stationary colored Gaussian with power spectrum \eqref{spectrum1}, while the minimum information leakage rate is given by \eqref{leakage1}.
\end{IEEEproof}

Note the lower bound is essentially given by the channel leakage of the virtual channel 
\begin{flalign}
\overline{\mathbf{x}}_{k} = \mathbf{x}_{k} + \mathbf{n}_{k}.
\end{flalign}

It can be verified that in general the more distortion allowed (i.e., the larger $D$ is), the less privacy leakage will occur (i.e., the smaller $L \left( S_{\mathbf{x}} \left( \omega \right), D \right)$ becomes). This privacy-distortion tradeoff is analytically captured in Theorem~\ref{privacy1}. In addition, when $\left\{ \mathbf{x}_{k} \right\}$ is further assumed to be stationary white Gaussian with variance $\sigma_{\mathbf{x}}^2$, the privacy-distortion tradeoff in Theorem~\ref{privacy1} reduces to a more explicit form as
\begin{flalign} 
\inf_{\mathbb{E} \left[ \left( \mathbf{x}_k - \overline{\mathbf{x}}_{k} \right)^2 \right] \leq D} I_{\infty} \left( \mathbf{x} ; \overline{\mathbf{x}} \right)
= L \left( \sigma_{\mathbf{x}}^2, D \right) 
= \frac{1}{2} \log \left( 1 + \frac{\sigma_{\mathbf{x}}^2}{D} \right),
\end{flalign} 
or equivalently,
\begin{flalign} 
D = \frac{\sigma_{\mathbf{x}}^2}{2^{2 L \left( \sigma_{\mathbf{x}}^2, D \right) } - 1}.
\end{flalign} 

It is also worth mentioning that this privacy-distortion tradeoff is
fundamentally different from the rate distortion function, which we will discuss in detail after Theorem~\ref{finitetime} in the sequel.

Note also that 
\begin{flalign}
I_{\infty} \left( \mathbf{x} ; \overline{\mathbf{x}} \right)
= h_{\infty} \left( \mathbf{x} \right) - h_{\infty} \left( \mathbf{x} | \overline{\mathbf{x}} \right),
\end{flalign}
and hence
\begin{flalign}
h_{\infty} \left( \mathbf{x} | \overline{\mathbf{x}} \right)
= h_{\infty} \left( \mathbf{x} \right) - I_{\infty} \left( \mathbf{x} ; \overline{\mathbf{x}} \right) 
= \frac{1}{2 \pi} \int_{0}^{2 \pi} \log \sqrt{ 2 \pi \mathrm{e}  S_{\mathbf{x}} \left( \omega \right) } \mathrm{d} \omega - I_{\infty} \left( \mathbf{x} ; \overline{\mathbf{x}} \right).
\end{flalign}
Since $S_{\mathbf{x}} \left( \omega \right)$ is pre-given, minimizing $I_{\infty} \left( \mathbf{x} ; \overline{\mathbf{x}} \right)$ is in fact equivalent to maximizing $h_{\infty} \left( \mathbf{x} | \overline{\mathbf{x}} \right)$, which is another privacy measure that is oftentimes employed in estimation problems (see, e.g., \cite{Cov:06, FangITW19} and the references therein). Particularly, it holds that
\begin{flalign}
\sup_{\mathbb{E} \left[ \left( \mathbf{x}_k - \overline{\mathbf{x}}_{k} \right)^2 \right] \leq D} h_{\infty} \left( \mathbf{x} | \overline{\mathbf{x}} \right) 
&= \frac{1}{2 \pi} \int_{0}^{2 \pi} \log \sqrt{ 2 \pi \mathrm{e}  S_{\mathbf{x}} \left( \omega \right) } \mathrm{d} \omega 
- \frac{1}{2 \pi} \int_{0}^{2 \pi} \log \sqrt{ 1 + \frac{S_{\mathbf{x}} \left( \omega \right)}{N \left( \omega \right)} } \mathrm{d} \omega \nonumber \\
&= \frac{1}{2 \pi} \int_{0}^{2 \pi} \log \sqrt{ 2 \pi \mathrm{e} \frac{ S_{\mathbf{x}} \left( \omega \right) N \left( \omega \right)}{ S_{\mathbf{x}} \left( \omega \right) + N \left( \omega \right)} }  \mathrm{d} \omega,
\end{flalign} 
where $N \left( \omega \right)$ is given by \eqref{spectrum1}.

On the other hand, the dual problem to that of Theorem~\ref{privacy1} would be: Given a certain privacy level, what is the minimum distortion between the original data and masked data? The following corollary answers this question.

\begin{corollary}
	Consider a data stream $\left\{ \mathbf{x}_{k} \right\}, \mathbf{x}_{k} \in \mathbb{R}$. Suppose that $\left\{ \mathbf{x}_{k} \right\}$ is stationary colored Gaussian with power spectrum $S_{\mathbf{x}} \left( \omega \right)$. For the sake of privacy, a noise $\left\{ \mathbf{n}_{k} \right\}, \mathbf{n}_{k} \in \mathbb{R}$, that is independent of $\left\{ \mathbf{x}_{k} \right\}$ is to be added to $\left\{ \mathbf{x}_{k} \right\}$ as its privacy mask, resulting in a masked streaming data $\left\{ \overline{\mathbf{x}}_{k} \right\}, \overline{\mathbf{x}}_{k} = \mathbf{x}_{k} + \mathbf{n}_{k}$,  whereas the properties of $\left\{ \mathbf{n}_{k} \right\}$ can be designed. Then, in order to make sure that the information leakage is upper bounded by a constant $R > 0$ as
	\begin{flalign}
	I_{\infty} \left( \mathbf{x} ; \overline{\mathbf{x}} \right) \leq R,
	\end{flalign}
	the minimum distortion between $\left\{ \mathbf{x}_{k} \right\}$ and $\left\{ \overline{\mathbf{x}}_{k} \right\}$ is given by
	\begin{flalign}
	\inf_{ I_{\infty} \left( \mathbf{x} ; \overline{\mathbf{x}} \right) \leq R } \mathbb{E} \left[ \left( \mathbf{x}_k - \overline{\mathbf{x}}_{k} \right)^2 \right] 
	= \frac{1}{2 \pi} \int_{0}^{2 \pi} \frac{\zeta}{2 \left[ 1 + \sqrt{1 + \frac{\zeta}{S_{\mathbf{x}} \left( \omega \right)}} \right]} \mathrm{d} \omega,
	\end{flalign} 
	where $\zeta \geq 0$ satisfies
	\begin{flalign}
	\frac{1}{2 \pi} \int_{0}^{2 \pi} \log \sqrt{ 1 + \frac{S_{\mathbf{x}} \left( \omega \right)}{\frac{\zeta}{2 \left[ 1 + \sqrt{1 + \frac{\zeta}{S_{\mathbf{x}} \left( \omega \right)}} \right]}} } \mathrm{d} \omega 
	= \frac{1}{2 \pi} \int_{0}^{2 \pi} \log \sqrt{ 1 + \frac{2}{\zeta} \left[ 1 + \sqrt{1 + \frac{\zeta}{S_{\mathbf{x}} \left( \omega \right)}} \right] S_{\mathbf{x}} \left( \omega \right) } \mathrm{d} \omega
	= R.
	\end{flalign} 
	Furthermore, in order to achieve this minimum distortion, the noise
	$\left\{ \mathbf{n}_{k} \right\}$ should be chosen as a stationary colored Gaussian process with power spectrum
	\begin{flalign}
	N \left( \omega \right) = \frac{\zeta}{2 \left[ 1 + \sqrt{1 + \frac{\zeta}{S_{\mathbf{x}} \left( \omega \right)}} \right]}.
	\end{flalign} 
\end{corollary}

Consider next the case of output power constraint.

\begin{theorem} \label{privacy2}
	Consider a data stream $\left\{ \mathbf{x}_{k} \right\}, \mathbf{x}_{k} \in \mathbb{R}$. Suppose that $\left\{ \mathbf{x}_{k} \right\}$ is stationary colored Gaussian with power spectrum $S_{\mathbf{x}} \left( \omega \right)$. For the sake of privacy, a noise $\left\{ \mathbf{n}_{k} \right\}, \mathbf{n}_{k} \in \mathbb{R}$, that is independent of $\left\{ \mathbf{x}_{k} \right\}$ is to be added to $\left\{ \mathbf{x}_{k} \right\}$ as its privacy mask, resulting in a masked streaming data $\left\{ \overline{\mathbf{x}}_{k} \right\}, \overline{\mathbf{x}}_{k} = \mathbf{x}_{k} + \mathbf{n}_{k}$,  whereas the properties of $\left\{ \mathbf{n}_{k} \right\}$ can be designed subject to a power constraint on the masked data $\left\{ \overline{\mathbf{x}}_{k} \right\}$ as $\mathbb{E} \left[ \overline{\mathbf{x}}_k^2 \right] \leq \overline{X}$.
	Then, in order to minimize the information leakage rate
	\begin{flalign}
	I_{\infty} \left( \mathbf{x} ; \overline{\mathbf{x}} \right),
	\end{flalign}
	the noise
	$\left\{ \mathbf{n}_{k} \right\}$ should be chosen as a stationary colored Gaussian process. In addition, the power spectrum of $\left\{ \mathbf{n}_{k} \right\}$ should be chosen as
	\begin{flalign} 
	N \left( \omega \right) = \frac{\zeta}{2 \left[ 1 + \sqrt{1 + \frac{\zeta}{S_{\mathbf{x}} \left( \omega \right)}} \right]},
	\end{flalign} 
	where $\zeta \geq 0$ satisfies
	\begin{flalign}
	\frac{1}{2 \pi} \int_{0}^{2 \pi} N \left( \omega \right) \mathrm{d} \omega 
	= \frac{1}{2 \pi} \int_{0}^{2 \pi} \frac{\zeta}{2 \left[ 1 + \sqrt{1 + \frac{\zeta}{S_{\mathbf{x}} \left( \omega \right)}} \right]} \mathrm{d} \omega 
	= \overline{X} - \frac{1}{2 \pi} \int_{0}^{2 \pi} S_{\mathbf{x}} \left( \omega \right) \mathrm{d} \omega.
	\end{flalign} 
	Correspondingly, the minimum information leakage rate is given by
	\begin{flalign}
	\inf_{\mathbb{E} \left[ \overline{\mathbf{x}}_k^2 \right] \leq \overline{X}} I_{\infty} \left( \mathbf{x} ; \overline{\mathbf{x}} \right)
	= L \left( S_{\mathbf{x}} \left( \omega \right), \overline{X} - \frac{1}{2 \pi} \int_{0}^{2 \pi} S_{\mathbf{x}} \left( \omega \right) \mathrm{d} \omega \right) 
	= \frac{1}{2 \pi} \int_{0}^{2 \pi} \log \sqrt{ 1 + \frac{S_{\mathbf{x}} \left( \omega \right)}{N \left( \omega \right)} } \mathrm{d} \omega.
	\end{flalign} 
\end{theorem}

\begin{IEEEproof}
	Since $\left\{ \mathbf{n}_{k} \right\}$ is independent of $\left\{ \mathbf{x}_{k} \right\}$, the power constraint reduces to
	\begin{flalign}
	\mathbb{E} \left[ \mathbf{n}_k^2 \right]
	\leq \overline{X} - \frac{1}{2 \pi} \int_{0}^{2 \pi} S_{\mathbf{x}} \left( \omega \right) \mathrm{d} \omega. \nonumber
	\end{flalign} 
	Then, Theorem~\ref{privacy2} follows by invoking Theorem~\ref{privacy1}.
\end{IEEEproof}

We may again consider the following dual problem.

\begin{corollary}
	Consider a data stream $\left\{ \mathbf{x}_{k} \right\}, \mathbf{x}_{k} \in \mathbb{R}$. Suppose that $\left\{ \mathbf{x}_{k} \right\}$ is stationary colored Gaussian with power spectrum $S_{\mathbf{x}} \left( \omega \right)$. For the sake of privacy, a noise $\left\{ \mathbf{n}_{k} \right\}, \mathbf{n}_{k} \in \mathbb{R}$, that is independent of $\left\{ \mathbf{x}_{k} \right\}$ is to be added to $\left\{ \mathbf{x}_{k} \right\}$ as its privacy mask, resulting in a masked streaming data $\left\{ \overline{\mathbf{x}}_{k} \right\}, \overline{\mathbf{x}}_{k} = \mathbf{x}_{k} + \mathbf{n}_{k}$,  whereas the properties of $\left\{ \mathbf{n}_{k} \right\}$ can be designed. Then, in order to make sure that the information leakage is upper bounded by a constant $R > 0$ as
	\begin{flalign}
	I_{\infty} \left( \mathbf{x} ; \overline{\mathbf{x}} \right) \leq R,
	\end{flalign}
	the minimum power of the masked data $\left\{ \overline{\mathbf{x}}_{k} \right\}$ is given by
	\begin{flalign}
	\inf_{ I_{\infty} \left( \mathbf{x} ; \overline{\mathbf{x}} \right) \leq R }\mathbb{E} \left[ \overline{\mathbf{x}}_{k}^2 \right] 
	= \frac{1}{2 \pi} \int_{0}^{2 \pi} \frac{\zeta}{2 \left[ 1 + \sqrt{1 + \frac{\zeta}{S_{\mathbf{x}} \left( \omega \right)}} \right]} \mathrm{d} \omega  + \frac{1}{2 \pi} \int_{0}^{2 \pi} S_{\mathbf{x}} \left( \omega \right) \mathrm{d} \omega,
	\end{flalign} 
	where $\zeta \geq 0$ satisfies
	\begin{flalign}
	\frac{1}{2 \pi} \int_{0}^{2 \pi} \log \sqrt{ 1 + \frac{2}{\zeta} \left[ 1 + \sqrt{1 + \frac{\zeta}{S_{\mathbf{x}} \left( \omega \right)}} \right] S_{\mathbf{x}} \left( \omega \right) } \mathrm{d} \omega
	= R.
	\end{flalign} 
	Furthermore, in order to achieve this minimum distortion, the noise
	$\left\{ \mathbf{n}_{k} \right\}$ should be chosen as a stationary colored Gaussian process with power spectrum
	\begin{flalign}
	N \left( \omega \right) = \frac{\zeta}{2 \left[ 1 + \sqrt{1 + \frac{\zeta}{S_{\mathbf{x}} \left( \omega \right)}} \right]}.
	\end{flalign} 
\end{corollary}  

\subsubsection{Auto-Regressive Moving Average (ARMA) Data}

Particularly, consider the case that the data stream $\left\{ \mathbf{x}_k \right\}, \mathbf{x}_k \in \mathbb{R}$ is the output of a linear time-invariant filter $F \left( z \right)$ with input $\left\{ \widehat{\mathbf{x}}_k \right\}, \widehat{\mathbf{x}}_k \in \mathbb{R}$, whereas $\left\{ \widehat{\mathbf{x}}_k \right\}$ is stationary white Gaussian with variance $\sigma_{\widehat{\mathbf{x}}}^2$. Particularly, when
\begin{flalign}
F \left( z \right) 
= \frac{ 1 + \sum_{i=1}^{p} f_{i} z^{-i} }{ 1 + \sum_{j=1}^{q} g_{j} z^{-j} },
\end{flalign}
which is assumed to be stable and minimum-phase, $\left\{ \mathbf{x}_k \right\}$ is an
ARMA data stream. In time domain, this is equivalent to
\begin{flalign} 
\mathbf{x}_{k} 
&= - \sum_{j=1}^{q} g_{j} \mathbf{x}_{k-j} + \widehat{\mathbf{x}}_k + \sum_{i=1}^{p} f_{i} \widehat{\mathbf{x}}_{k-i}.
\end{flalign}
In addition, the power spectrum of $\left\{ \mathbf{x}_k \right\}$ is given by 
\begin{flalign}
S_{\mathbf{x}} \left( \omega \right)
= \left| F \left( \mathrm{e}^{\mathrm{j} \omega} \right) \right|^2 \sigma_{\widehat{\mathbf{x}}}^2
= \left| \frac{ 1 + \sum_{i=1}^{p} f_{i} \mathrm{e}^{- i \mathrm{j} \omega} }{ 1 + \sum_{j=1}^{q} g_{j} \mathrm{e}^{- j \mathrm{j} \omega} } \right|^2 \sigma_{\widehat{\mathbf{x}}}^2.
\end{flalign}


\subsection{Non-Stationary Data Stream}

Let us proceed to consider the non-stationary data streams in this subsection. We consider the distortion constraint first.

\begin{theorem} \label{privacy11}
	Consider a data stream $\left\{ \mathbf{x}_{k} \right\}, \mathbf{x}_{k} \in \mathbb{R}$. Suppose that $ \mathbf{x}_{k} = \mathbf{h}_{k} \widehat{\mathbf{x}}_{k}$, where $\left\{ \widehat{\mathbf{x}}_{k} \right\}, \widehat{\mathbf{x}}_k \in \mathbb{R}$, is stationary white Gaussian with variance $\sigma_{\widehat{\mathbf{x}}}^2$, and $\left\{ \mathbf{h}_k \right\}, \mathbf{h}_k \in \mathbb{R}$, is the fading process. Suppose that the fading time of $\left\{ \mathbf{h}_k \right\}$ is relatively small compared to transmission duration. (Note that $\left\{ \mathbf{x}_{k} \right\}$ is no longer stationary, since the variance $\sigma_{\mathbf{x}_{k}}^2 = \left| h_{k} \right|^2 \sigma_{\widehat{\mathbf{x}}}^2 $ of $\mathbf{x}_{k}$ is in general time varying, as a function of $ h_{k}$ which denotes the realization of $ \mathbf{h}_{k}$.)
    For the sake of privacy, a noise $\left\{ \mathbf{n}_{k} \right\}, \mathbf{n}_{k} \in \mathbb{R}$, that is independent of $\left\{ \mathbf{x}_{k} \right\}$ and $\left\{ \mathbf{h}_{k} \right\}$ is to be added to $\left\{ \mathbf{x}_{k} \right\}$ as its privacy mask, resulting in a masked streaming data $\left\{ \overline{\mathbf{x}}_{k} \right\}, \overline{\mathbf{x}}_{k} = \mathbf{x}_{k} + \mathbf{n}_{k}
	= \mathbf{h}_{k} \widehat{\mathbf{x}}_{k} + \mathbf{n}_{k}$,  whereas the properties of $\left\{ \mathbf{n}_{k} \right\}$ can be designed subject to a distortion constraint 
	\begin{flalign}
	\mathbb{E} \left[ \left( \mathbf{x}_k - \overline{\mathbf{x}}_{k} \right)^2 \right] \leq D.
	\end{flalign} 
	Assume further that $\left\{ \widehat{\mathbf{x}}_{k} \right\}$, $\left\{ \mathbf{h}_{k} \right\}$, and $\left\{ \mathbf{n}_{k} \right\}$ are mutually independent.
	
	{\em Case~1: Without noise side information.} In order to minimize the  information leakage rate
	\begin{flalign}
	I_{\infty} \left( \mathbf{x} ; \overline{\mathbf{x}} \right),
	\end{flalign}
	the noise
	$\left\{ \mathbf{n}_{k} \right\}$ should be chosen as a stationary white Gaussian process with variance $D$. Correspondingly, the minimum information leakage rate is given by
	\begin{flalign}
	\inf_{ \mathbb{E} \left[ \left( \mathbf{x}_k - \overline{\mathbf{x}}_{k} \right)^2 \right] \leq D } I_{\infty} \left( \mathbf{x} ; \overline{\mathbf{x}} \right)
	=\mathbb{E} \left[\frac{1}{2} \log \left( 1+\frac{\left| \mathbf{h}_k \right|^{2} \sigma_{\mathbf{x}}^2 }{D} \right) \right].
	\end{flalign}
	
	{\em Case~2: With noise side information.} 
	In order to minimize the  information leakage rate
	\begin{flalign}
	I_{\infty} \left( \mathbf{x} ; \overline{\mathbf{x}} \right),
	\end{flalign}
	the noise
	$\left\{ \mathbf{n}_{k} \right\}$ should be chosen as a non-stationary white Gaussian process. Furthermore, the (time-varying) variance of $ \mathbf{n}_{k} $  should be chosen as
	\begin{flalign}
	N_{k} = \frac{\zeta}{2\left(\sqrt{1+\frac{\zeta }{\left| h_k \right|^2 \sigma_{\mathbf{x}}^2}}+1
		\right)}, 
	\end{flalign}
	where $\zeta \geq 0$ satisfies
	\begin{flalign}
	\mathbb{E} \left[   N_{k} \right]
	= \mathbb{E} \left[   \frac{\zeta}{2\left(\sqrt{1+\frac{\zeta }{\left| \mathbf{h}_k \right|^2 \sigma_{\mathbf{x}}^2}}+1
		\right)} \right]
	=D.
	\end{flalign}
	Moreover, the minimum information leakage rate is given by
	\begin{flalign}
	\inf_{\mathbb{E} \left[ \left( \mathbf{x}_k - \overline{\mathbf{x}}_{k} \right)^2 \right] \leq D} I_{\infty} \left( \mathbf{x} ; \overline{\mathbf{x}} \right) = \mathbb{E} \left[ \frac{1}{2}\log
	\left(1+\frac{ \left| \mathbf{h}_k \right|^2 \sigma_{\mathbf{x}}^2}{N_{k}}\right) \right].
	\end{flalign}
\end{theorem}

\begin{IEEEproof}
	Note that the distortion constraint $
	\mathbb{E} \left[ \left( \mathbf{x}_k - \overline{\mathbf{x}}_{k} \right)^2 \right] \leq D $
	is equivalent to being with a power constraint $\mathbb{E} \left[ \mathbf{n}_k^2 \right] \leq D$, since $\overline{\mathbf{x}}_{k} = \mathbf{x}_{k} + \mathbf{n}_k $ and thus $\left( \mathbf{x}_k - \overline{\mathbf{x}}_{k} \right)^2 = \mathbf{n}_k^2$.
	The rest of the proof proceeds as in the that of Theorem~\ref{fading}, by viewing $\mathbf{n}_{k}$ and $\overline{\mathbf{x}}_{k}$ as $\mathbf{z}_{k}$ and $\mathbf{y}_{k}$ therein, respectively. 
\end{IEEEproof}

The dual problem would be: Given a certain privacy level, what is the minimum distortion? The following corollary answers this question.

\begin{corollary}
	Consider a data stream $\left\{ \mathbf{x}_{k} \right\}, \mathbf{x}_{k} \in \mathbb{R}$. Suppose that $ \mathbf{x}_{k} = \mathbf{h}_{k} \widehat{\mathbf{x}}_{k}$, where $\left\{ \widehat{\mathbf{x}}_{k} \right\}, \widehat{\mathbf{x}}_k \in \mathbb{R}$, is stationary white Gaussian with variance $\sigma_{\widehat{\mathbf{x}}}^2$, and $\left\{ \mathbf{h}_k \right\}, \mathbf{h}_k \in \mathbb{R}$, is the fading process. Suppose that the fading time of $\left\{ \mathbf{h}_k \right\}$ is relatively small compared to transmission duration. (Note that $\left\{ \mathbf{x}_{k} \right\}$ is no longer stationary, since the variance $\sigma_{\mathbf{x}_{k}}^2 = \left| h_{k} \right|^2 \sigma_{\widehat{\mathbf{x}}}^2 $ of $\mathbf{x}_{k}$ is in general time varying, as a function of $ h_{k}$ which denotes the realization of $ \mathbf{h}_{k}$.)
	For the sake of privacy, a noise $\left\{ \mathbf{n}_{k} \right\}, \mathbf{n}_{k} \in \mathbb{R}$, that is independent of $\left\{ \mathbf{x}_{k} \right\}$ and $\left\{ \mathbf{h}_{k} \right\}$ is to be added to $\left\{ \mathbf{x}_{k} \right\}$ as its privacy mask, resulting in a masked streaming data $\left\{ \overline{\mathbf{x}}_{k} \right\}, \overline{\mathbf{x}}_{k} = \mathbf{x}_{k} + \mathbf{n}_{k}
	= \mathbf{h}_{k} \widehat{\mathbf{x}}_{k} + \mathbf{n}_{k}$,  whereas the properties of $\left\{ \mathbf{n}_{k} \right\}$ can be designed.
	Assume further that $\left\{ \widehat{\mathbf{x}}_{k} \right\}$, $\left\{ \mathbf{h}_{k} \right\}$, and $\left\{ \mathbf{n}_{k} \right\}$ are mutually independent.
	
	{\em Case~1: Without noise side information.} 
	In order to make sure that the information leakage is upper bounded by a constant $R > 0$ as
	\begin{flalign}
	I_{\infty} \left( \mathbf{x} ; \overline{\mathbf{x}} \right) \leq R,
	\end{flalign}
	the minimum distortion between $\left\{ \mathbf{x}_{k} \right\}$ and $\left\{ \overline{\mathbf{x}}_{k} \right\}$ is given by
	\begin{flalign}
	&\inf_{ I_{\infty} \left( \mathbf{x} ; \overline{\mathbf{x}} \right) \leq R } \mathbb{E} \left[ \left( \mathbf{x}_k - \overline{\mathbf{x}}_{k} \right)^2 \right]
	= D,
	\end{flalign} 
	where $D > 0$ satisfies
	\begin{flalign}
	\mathbb{E} \left[\frac{1}{2} \log \left( 1+\frac{\left| \mathbf{h}_k \right|^{2} \sigma_{\mathbf{x}}^2 }{D} \right) \right] = R.
	\end{flalign} 
	Furthermore, in order to achieve this minimum distortion, the noise
	$\left\{ \mathbf{n}_{k} \right\}$ should be chosen as a stationary white Gaussian process with variance $D$.
	
	{\em Case~2: With noise side information.} 
	In order to make sure that the information leakage is upper bounded by a constant $R > 0$ as
	\begin{flalign}
	I_{\infty} \left( \mathbf{x} ; \overline{\mathbf{x}} \right) \leq R,
	\end{flalign}
	the minimum distortion between $\left\{ \mathbf{x}_{k} \right\}$ and $\left\{ \overline{\mathbf{x}}_{k} \right\}$ is given by
	\begin{flalign}
	&\inf_{ I_{\infty} \left( \mathbf{x} ; \overline{\mathbf{x}} \right) \leq R } \mathbb{E} \left[ \left( \mathbf{x}_k - \overline{\mathbf{x}}_{k} \right)^2 \right]
	= \mathbb{E} \left[   \frac{\zeta}{2\left(\sqrt{1+\frac{\zeta }{\left| \mathbf{h}_k \right|^2 \sigma_{\mathbf{x}}^2}}+1
		\right)} \right],
	\end{flalign} 
	where $\zeta \geq 0$ satisfies
	\begin{flalign}
	\mathbb{E} \left[ \frac{1}{2}\log
	\left(1+\frac{ \left| \mathbf{h}_k \right|^2 \sigma_{\mathbf{x}}^2}{\frac{\zeta}{2\left(\sqrt{1+\frac{\zeta }{\left| \mathbf{h}_k \right|^2 \sigma_{\mathbf{x}}^2}}+1
			\right)} }\right) \right]
		= \mathbb{E} \left[ \frac{1}{2}\log
	\left( 1 + \frac{2}{\zeta} \left[ 1 + \sqrt{1 + \frac{\zeta}{\left| \mathbf{h}_k \right|^2 \sigma_{\mathbf{x}}^2 }} \right] \left| \mathbf{h}_k \right|^2 \sigma_{\mathbf{x}}^2 \right)  \right] = R.
	\end{flalign} 
	Furthermore, in order to achieve this minimum distortion, the noise
	$\left\{ \mathbf{n}_{k} \right\}$ should be chosen as a non-stationary white Gaussian process, whereas the variance of $\mathbf{n}_{k}$ should be chosen as
	\begin{flalign}
	N_{k} = \frac{\zeta}{2\left(\sqrt{1+\frac{\zeta }{\left| h_k \right|^2 \sigma_{\mathbf{x}}^2}}+1
		\right)},
	\end{flalign}
	which is time varying.
\end{corollary}

Similarly to the distortion constraint case, the case of output power constraint and its dual problem may be analyzed similarly based upon Theorem~\ref{output2}.

\subsection{Finite-Time Data Stream}

This subsection is devoted to finite-time data streams.
We first consider the case of distortion constraint.

\begin{theorem} \label{finitetime}
	Consider a finite-time data stream $\mathbf{x}_{k}  \in \mathbb{R}, k = 0, \ldots, K$. Suppose that $ \mathbf{x}_{0, \ldots, K} $ is Gaussian with covariance $\Sigma_{\mathbf{x}_{0, \ldots, K}}$. 
	Suppose that the eigen-decomposition of $\Sigma_{\mathbf{x}_{0,\ldots,K}}$ be given by
	\begin{flalign}
	\Sigma_{\mathbf{x}_{0,\ldots,K}}
	=U_{\mathbf{x}_{0,\ldots,K}}\Lambda_{\mathbf{x}_{0,\ldots,K}}U^{\mathrm{T}}_{\mathbf{x}_{0,\ldots,K}}, 
	\end{flalign} 
	where
	\begin{flalign}
	\Lambda_{\mathbf{x}_{0,\ldots,K}}
	=\mathrm{diag} \left( \lambda_{0},\ldots,\lambda_{K} \right). 
	\end{flalign}
	For the sake of privacy, a noise $\mathbf{n}_{k} \in \mathbb{R}$ that is independent of $\mathbf{x}_{0, \ldots, K}$ is to be added to $\mathbf{x}_{k}$ at each time $k = 0, \ldots, K$ as its privacy mask, resulting in a masked data $ \overline{\mathbf{x}}_{0, \ldots, K}$ with $ \overline{\mathbf{x}}_{k} = \mathbf{x}_{k} + \mathbf{n}_{k}, k = 0, \ldots, K$. The properties of $\mathbf{n}_{0, \ldots, K}$ can be designed subject to a total distortion constraint 
	\begin{flalign}
	\sum_{k=0}^{K} \mathbb{E} \left[ \left( \mathbf{x}_k - \overline{\mathbf{x}}_{k} \right)^2 \right] \leq D.
	\end{flalign} 
	Then, in order to minimize the total information leakage 
	\begin{flalign}
	I \left( \mathbf{x}_{0, \ldots, K} ; \overline{\mathbf{x}}_{0, \ldots, K} \right),
	\end{flalign}
	the noise
	$\mathbf{n}_{0, \ldots, K}$ should be chosen as a Gaussian sequence with covariance 
	\begin{flalign}
	\Sigma_{\mathbf{n}_{0,\ldots,K}}
	=U_{\mathbf{x}_{0,\ldots,K}} 
	\mathrm{diag} \left( N_{0},\ldots,N_{K} \right)
	U^{\mathrm{T}}_{\mathbf{x}_{0,\ldots,K}}.
	\end{flalign} 
	Herein, $N_{k}$ is given by
	\begin{flalign}
	N_{k} = \frac{\zeta}{2 \left( 1 + \sqrt{1 + \frac{\zeta}{\lambda_{k}}} \right)},~k = 0, \ldots, K,
	\end{flalign} 
	where $\zeta \geq 0$ satisfies
	\begin{flalign} 
	\sum_{k=0}^{K} N_{k}
	= \sum_{k=0}^{K} \frac{\zeta}{2 \left( 1 + \sqrt{1 + \frac{\zeta}{\lambda_{k}}} \right)}
	= D.
	\end{flalign} 
	Correspondingly, the minimum total information leakage is
	\begin{flalign} 
	\inf_{\sum_{k=0}^{K}  \mathbb{E} \left[ \left( \mathbf{x}_k - \overline{\mathbf{x}}_{k} \right)^2 \right] \leq D} I \left( \mathbf{x}_{0,\ldots,K} ; \overline{\mathbf{x}}_{0,\ldots,K} \right)
	= \sum_{k=0}^{K} \frac{1}{2} \log \left( 1 + \frac{\lambda_{k}}{N_{k}} \right).
	\end{flalign} 
\end{theorem}

\begin{IEEEproof}
	The proof resembles the first half of the proof for Theorem~\ref{ACGI}, by viewing $\mathbf{n}_{k}$ and $\overline{\mathbf{x}}_{k}$ as $\mathbf{z}_{k}$ and $\mathbf{y}_{k}$ therein, respectively. 
	The basic idea is still to consider the $k+1$ consecutive uses of this channel 
	as $k+1$ channels in parallel \cite{Cov:06}. On the other hand, the distortion constraint \begin{flalign} \sum_{k=0}^{K}
	\mathbb{E} \left[ \left( \mathbf{x}_k - \overline{\mathbf{x}}_{k} \right)^2 \right] \leq D \nonumber
	\end{flalign} 
	is equivalent to being with a power constraint 
	\begin{flalign} 	\sum_{k=0}^{K} \mathbb{E} \left[ \mathbf{n}_k^2 \right] \leq D,\nonumber
	\end{flalign} 
	since $\overline{\mathbf{x}}_{k} = \mathbf{x}_{k} + \mathbf{n}_k $ and thus $\left( \mathbf{x}_k - \overline{\mathbf{x}}_{k} \right)^2 = \mathbf{n}_k^2$.
\end{IEEEproof}

Consider now the special case when $\Sigma_{\mathbf{x}_{0,\ldots,K}}$ is diagonal with 
\begin{flalign}
\Sigma_{\mathbf{x}_{0,\ldots,K}}
= \mathrm{diag} \left( \sigma_{\mathbf{x}_{0}}^2, \ldots, \sigma_{\mathbf{x}_{K}}^2 \right), 
\end{flalign}
where $\sigma_{\mathbf{x}_{k}}^2$ denotes the variance of $\mathbf{x}_{k}$ for $k = 0, 
\ldots, K$. In this case, $N_{k}$ is equivalent to the distortion of $\mathbf{x}_{k}$ as 
\begin{flalign}
N_{k} = \mathbb{E} \left[ \left( \mathbf{x}_k - \overline{\mathbf{x}}_{k} \right)^2 \right] = D_{k}.
\end{flalign}
Accordingly, Theorem~\ref{finitetime} reduces to
\begin{flalign} 
\inf_{\sum_{k=0}^{K}  \mathbb{E} \left[ \left( \mathbf{x}_k - \overline{\mathbf{x}}_{k} \right)^2 \right] \leq D} I \left( \mathbf{x}_{0,\ldots,K} ; \overline{\mathbf{x}}_{0,\ldots,K} \right) 
= \sum_{k=0}^{K} \frac{1}{2} \log \left( 1 + \frac{\sigma_{\mathbf{x}_{k}}^2}{D_{k}} \right),
\end{flalign}  
where
\begin{flalign}
D_{k} = \frac{\zeta}{2 \left( 1 + \sqrt{1 + \frac{\zeta}{\sigma_{\mathbf{x}_{k}}^2}} \right)},~k = 0, \ldots, K,
\end{flalign} 
and $\zeta \geq 0$ satisfies
\begin{flalign} 
\sum_{k=0}^{K} D_{k}
= \sum_{k=0}^{K} \frac{\zeta}{2 \left( 1 + \sqrt{1 + \frac{\zeta}{\sigma_{\mathbf{x}_{k}}^2}} \right)}
= D.
\end{flalign} 
In addition to we discussed earlier after Corollary~\ref{ACGIc}, it is once again manifested that this privacy-distortion tradeoff is fundamentally different from the rate distortion function. Note that one essential difference is that herein $ \mathbf{x}_k - \overline{\mathbf{x}}_{k}$ (and accordingly, $\mathbf{n}_k$ in channel leakage) is independent of $\mathbf{x}_k$, while for the rate distortion function, $ \mathbf{x}_k - \widehat{\mathbf{x}}_{k}$ is not necessarily independent of $\mathbf{x}_k$. We will, however, leave detailed discussions on this topic to future research.

The dual problem is analyzed as follows.

\begin{corollary}
	Consider a finite-time data stream $\mathbf{x}_{k}  \in \mathbb{R}, k = 0, \ldots, K$. Suppose that $ \mathbf{x}_{0, \ldots, K} $ is Gaussian with covariance $\Sigma_{\mathbf{x}_{0, \ldots, K}}$. 
	Suppose that the eigen-decomposition of $\Sigma_{\mathbf{x}_{0,\ldots,K}}$ be given by
	\begin{flalign}
	\Sigma_{\mathbf{x}_{0,\ldots,K}}
	=U_{\mathbf{x}_{0,\ldots,K}}\Lambda_{\mathbf{x}_{0,\ldots,K}}U^{\mathrm{T}}_{\mathbf{x}_{0,\ldots,K}}, 
	\end{flalign} 
	where
	\begin{flalign}
	\Lambda_{\mathbf{x}_{0,\ldots,K}}
	=\mathrm{diag} \left( \lambda_{0},\ldots,\lambda_{K} \right). 
	\end{flalign}
	For the sake of privacy, a noise $\mathbf{n}_{k} \in \mathbb{R}$  that is independent of $\mathbf{x}_{0, \ldots, K}$ is to be added to $\mathbf{x}_{k}$ at each time $k = 0, \ldots, K$ as its privacy mask, resulting in a masked data $ \overline{\mathbf{x}}_{0, \ldots, K}$ with $ \overline{\mathbf{x}}_{k} = \mathbf{x}_{k} + \mathbf{n}_{k}, k = 0, \ldots, K$. The properties of $\mathbf{n}_{0, \ldots, K}$ can be designed. Then, in order to make sure that the total information leakage is upper bounded by a constant $R > 0$ as
	\begin{flalign}
	I \left( \mathbf{x}_{0, \ldots, K} ; \overline{\mathbf{x}}_{0, \ldots, K} \right) \leq R,
	\end{flalign}
	the minimum total distortion between $\mathbf{x}_{0, \ldots, K}$ and $\overline{\mathbf{x}}_{0, \ldots, K}$ is given by
	\begin{flalign}
	\inf_{ I \left( \mathbf{x}_{0, \ldots, K} ; \overline{\mathbf{x}}_{0, \ldots, K} \right) \leq R } \sum_{k=1}^{K} \mathbb{E} \left[ \left( \mathbf{x}_k - \overline{\mathbf{x}}_{k} \right)^2 \right] 
	= \sum_{k=0}^{K} \frac{\zeta}{2 \left( 1 + \sqrt{1 + \frac{\zeta}{\lambda_{k}}} \right)},
	\end{flalign} 
	where $\zeta \geq 0$ satisfies
	\begin{flalign}
	\sum_{k=0}^{K} \frac{1}{2} \log \left[ 1 + \frac{\lambda_{k}}{\frac{\zeta}{2 \left( 1 + \sqrt{1 + \frac{\zeta}{\lambda_{k}}} \right)}} \right] 
	= \sum_{k=0}^{K} \frac{1}{2} \log \left[ 1 + \frac{2}{\zeta} \left( 1 + \sqrt{1 + \frac{\zeta}{\lambda_{k}}} \right) \lambda_{k} \right] 
	= R.
	\end{flalign} 
	Furthermore, in order to achieve this minimum total distortion, the noise
	$\mathbf{n}_{0, \ldots, K}$ should be chosen as a Gaussian sequence with covariance 
	\begin{flalign}
	\Sigma_{\mathbf{n}_{0,\ldots,K}}
	=U_{\mathbf{x}_{0,\ldots,K}} 
	\mathrm{diag} \left( N_{0},\ldots,N_{K} \right)
	U^{\mathrm{T}}_{\mathbf{x}_{0,\ldots,K}},
	\end{flalign} 
	where $N_{k}$ is given by
	\begin{flalign}
	N_{k} = \frac{\zeta}{2 \left( 1 + \sqrt{1 + \frac{\zeta}{\lambda_{k}}} \right)},~k = 0, \ldots, K.
	\end{flalign} 
\end{corollary}  

Again, we may also analyze the case of output power constraint and its dual problem.

\section{Conclusion}

In this paper, we have formally introduced the notion of channel leakage as the minimum mutual information rate between the channel input and channel output, which characterizes the minimum information leakage rate to the malicious receiver.
We have obtained explicit formulas of channel leakage for the white Gaussian case, the colored Gaussian case, and the fading case. We have also derived analytical equations for the fundamental privacy-distortion tradeoffs and privacy-power tradeoffs in streaming data, including the stationary case, the non-stationary case, and the finite-time case, based upon the notion of channel leakage, together with explicit solutions for the optimal privacy mask design. Potential future research directions include the investigations of non-Gaussian data streams. It might also be interesting to see whether (and if yes, how) feedback in general can help decrease the channel leakage and thus improve the privacy tradeoffs.

\appendices

\section{Proof of Theorem~\ref{AWGI}} \label{appendix1}
Since $\left\{ \mathbf{x}_{k} \right\}$ is white, we have
\begin{flalign}
I \left(\mathbf{x}_{0,\ldots,k}; \mathbf{y}_{0,\ldots,k} \right)
&= h \left(\mathbf{x}_{0,\ldots,k} \right)-h \left(\mathbf{x}_{0,\ldots,k} \middle| \mathbf{y}_{0,\ldots,k} \right) 
= \sum_{i=0}^k h \left(\mathbf{x}_i\right)-h \left(\mathbf{z}_{0,\ldots,k} \middle| \mathbf{y}_{0,\ldots,k} \right) \nonumber \\
& = \sum_{i=0}^k h \left(\mathbf{x}_i\right)-\sum_{i=0}^k h \left(\mathbf{z}_{i} \middle| \mathbf{y}_{0,\ldots,k},\mathbf{z}_{0,\ldots,i-1} \right) 
\geq \sum_{i=0}^k h \left(\mathbf{x}_i\right)-\sum_{i=0}^k h \left(\mathbf{z}_{i} \middle| \mathbf{y}_{i}\right)\nonumber \\
&= \sum_{i=0}^k h \left(\mathbf{x}_i\right)-\sum_{i=0}^k h \left(\mathbf{x}_{i} \middle| \mathbf{y}_{i}\right) 
= \sum_{i=0}^k I \left(\mathbf{x}_i;\mathbf{y}_i\right),
\nonumber
\end{flalign}
and
\begin{flalign}
I \left(\mathbf{x}_{0,\ldots,k}; \mathbf{y}_{0,\ldots,k} \right)
= \sum_{i=0}^k I \left(\mathbf{x}_i;\mathbf{y}_i\right)
\nonumber
\end{flalign}
if $\left\{ \mathbf{z}_{k} \right\}$ is white. On the other hand, since $\mathbf{x}_i$ and $\mathbf{z}_i$ are independent, we have
\begin{flalign}
I\left(\mathbf{x}_i;\mathbf{y}_i \right)
=h\left(\mathbf{y}_i\right)-h\left(\mathbf{y}_i \middle| \mathbf{x}_i\right)
=h\left(\mathbf{y}_i\right)-h\left(\mathbf{z}_i \middle| \mathbf{x}_i\right) =h\left(\mathbf{y}_i\right)-h\left(\mathbf{z}_i\right), \nonumber
\end{flalign}
and
\begin{flalign}
I\left(\mathbf{z}_i;\mathbf{y}_i \right)
=h\left(\mathbf{y}_i\right)-h\left(\mathbf{y}_i \middle| \mathbf{z}_i\right)
=h\left(\mathbf{y}_i\right)-h\left(\mathbf{x}_i \middle| \mathbf{z}_i\right) =h\left(\mathbf{y}_i\right)-h\left(\mathbf{x}_i\right). \nonumber
\end{flalign}
Then, according to the entropy power inequality \cite{Cov:06}, we have
\begin{flalign}
&2^{2h\left(\mathbf{y}_i\right)}\geq 2^{2h\left(\mathbf{z}_i\right)}+2^{2h\left(\mathbf{x}_i\right)}, \nonumber
\end{flalign}
and hence
\begin{flalign}
2^{2\left[h\left(\mathbf{z}_i\right)-h\left(\mathbf{y}_i\right)\right]}+2^{2\left[h\left(\mathbf{x}_i\right)-h\left(\mathbf{y}_i\right)\right]}
\leq 1. \nonumber
\end{flalign}
Consequently,
\begin{flalign}
I \left(\mathbf{x}_i; \mathbf{y}_i \right)
= h\left(\mathbf{y}_i\right)-h\left(\mathbf{z}_i\right)
\geq -\frac{1}{2} \log \left\{ 1-2^{2\left[h\left(\mathbf{x}_i\right)-h\left(\mathbf{y}_i\right)\right]} \right\} 
=-\frac{1}{2} \log \left[ 1-2^{-2 I \left(\mathbf{z}_i; \mathbf{y}_i \right)} \right].
\nonumber
\end{flalign}
On the other hand, it can be verified \cite{Cov:06} that, subjecting to the power constraint $\mathbb{E} \left[ \mathbf{z}_i^{2} \right] \leq N$,  $I\left(\mathbf{z}_i;\mathbf{y}_i \right)$ reaches its maximum
\begin{flalign}
\frac{1}{2}\log \left( 1+\frac{N}{\sigma_{\mathbf{x}}^2} \right)
\nonumber
\end{flalign}
when $\mathbf{z}_i$ is Gaussian and $\mathbb{E} \left[ \mathbf{z}_i^{2} \right]=N$. Note also that if $\mathbf{z}_i$ is Gaussian, then
\begin{flalign}
2^{2h\left(\mathbf{y}_i\right)}
= 2^{2h\left(\mathbf{z}_i\right)}+2^{2h\left(\mathbf{x}_i\right)}, \nonumber
\end{flalign}
and thus
\begin{flalign}
I\left(\mathbf{x}_i; \mathbf{y}_i \right)=-\frac{1}{2} \log \left[ 1-2^{-2 I \left(\mathbf{z}_i; \mathbf{y}_i \right)} \right].  \nonumber
\end{flalign}
That is to say, subjecting to the power constraint $\mathbb{E} \left[ \mathbf{z}_i^{2} \right] \leq N$, 
$I\left(\mathbf{x}_i; \mathbf{y}_i \right)$ reaches its minimum if $\mathbf{z}_i$ is Gaussian and $\mathbb{E} \left[ \mathbf{z}_i^{2} \right]=N$, and the minimum is given by
\begin{flalign}
\min_{\mathbb{E} \left[ \mathbf{z}_i^{2} \right]\leq N} I \left(\mathbf{x}_i; \mathbf{y}_i \right)
=-\frac{1}{2} \log \left[ 1-2^{-\log \left( 1+\frac{N}{\sigma_{\mathbf{x}}^2} \right)} \right] 
=\frac{1}{2}\log \left( 1+\frac{\sigma_{\mathbf{x}}^2}{N} \right).  \nonumber
\end{flalign} 
As such, as $k \to \infty$, $\left\{ \mathbf{x}_{k} \right\}$, $\left\{ \mathbf{z}_{k} \right\}$, and $\left\{ \mathbf{y}_{k} \right\}$ are stationary white (cf. the proof of Theorem~\ref{ACGI}), and
\begin{flalign}
\min_{\mathbb{E} \left[ \mathbf{z}_i^{2} \right]\leq N} I \left(\mathbf{x}_i; \mathbf{y}_i \right),~\forall i \in \mathbb{N} 
&= \inf_{\mathbb{E} \left[ \mathbf{z}_k^{2} \right]\leq N} \lim_{k\to \infty} \frac{\sum_{i=0}^k I \left(\mathbf{x}_i;\mathbf{y}_i\right)}{k+1} 
= \inf_{\mathbb{E} \left[ \mathbf{z}_k^{2} \right]\leq N} \limsup_{k\to \infty} \frac{\sum_{i=0}^k I \left(\mathbf{x}_i;\mathbf{y}_i\right)}{k+1} \nonumber \\ 
& = \inf_{\mathbb{E} \left[ \mathbf{z}_k^{2} \right]\leq N} \limsup_{k \to \infty} \frac{I \left( \mathbf{x}_{0, \ldots, k}; \mathbf{y}_{0, \ldots, k} \right)}{k+1} 
= \inf_{\mathbb{E} \left[ \mathbf{z}_k^{2} \right]\leq N} I_{\infty} \left( \mathbf{x}; \mathbf{y} \right) 
= L. \nonumber
\end{flalign}
In other words,
\begin{flalign}
L=	\frac{1}{2}\log \left( 1+\frac{\sigma_{\mathbf{x}}^2}{N} \right),
\nonumber
\end{flalign}
which is achieved when $\left\{ \mathbf{z}_i \right\}$ is stationary white Gaussian with variance $\mathbb{E} \left[ \mathbf{z}_i^{2} \right]=N$.
Note that although mathematically this proof is in the same spirit as that for channel blurredness in \cite{fang2017towards} in general, the implications are totally different. To prevent possible confusion and for the sake of self-containment, in this paper we still present a step-by-step proof. This statement applies to the proofs of Theorem~\ref{ACGI} and Theorem~\ref{fading} (for certain steps) as well.

\section{Proof of Theorem~\ref{ACGI}}

We first consider the case of a finite number of parallel (dependent) channels with
\begin{flalign}
\mathbf{y} = \mathbf{x} + \mathbf{z}, \nonumber
\end{flalign}
where $ \mathbf{x},\mathbf{y},\mathbf{z} \in  \mathbb{R}^m$, and $ \mathbf{z} $ is independent of $ \mathbf{x} $. In addition, $\mathbf{x}$ is Gaussian with covariance $\Sigma_{\mathbf{x}}$, and the noise power constraint is given by 
\begin{flalign} 
\mathrm{tr} \left(\Sigma_{\mathbf{z} }  \right)
= \mathbb{E} \left[ \sum_{i=1}^{m}
\mathbf{z}^{2}\left( i \right) \right] \leq N. \nonumber
\end{flalign}
where $\mathbf{z} \left( i \right)$ denotes the $i$-th element of $\mathbf{z}$.
(Note that the case of parallel independent channels, as discussed in Corollary~\ref{ACGIc}, is a special case of that of dependent channels for when $\Sigma_{\mathbf{x}}$ is diagonal.)
In addition, since $\mathbf{x}$ and $ \mathbf{z}$ are independent, we have
\begin{flalign}
I\left(\mathbf{x} ;\mathbf{y} \right)
=h\left(\mathbf{y} \right)-h\left(\mathbf{y} \middle| \mathbf{x} \right)
=h\left(\mathbf{y} \right)-h\left(\mathbf{z} \middle| \mathbf{x} \right) =h\left(\mathbf{y} \right)-h\left(\mathbf{z} \right), \nonumber
\end{flalign}
and \begin{flalign}
\Sigma_{\mathbf{y}}
=\Sigma_{\mathbf{z} +\mathbf{x} }
=\Sigma_{\mathbf{z}} +\Sigma_{\mathbf{x}}. \nonumber
\end{flalign}
On the other hand, the minimum of $I \left(\mathbf{x}; \mathbf{y} \right)$ is achieved if $\mathbf{z}$ is Gaussian (see Section~11.9 of \cite{yeung2008information}), whereas when $\mathbf{z}$ is Gaussian, we have
\begin{flalign}
I \left(\mathbf{x}; \mathbf{y} \right)
&=h\left( \mathbf{y}\right)-h\left(\mathbf{z}\right)
=\frac{1}{2}\log \left[ \left(2\pi \mathrm{e} \right)^m  \det  \Sigma_{\mathbf{y}} \right]-\frac{1}{2}\log \left[ \left(2\pi \mathrm{e} \right)^m  \det  \Sigma_{\mathbf{z}}\right]
=\frac{1}{2}\log \frac{ \det  \Sigma_{\mathbf{y}}}{ \det  \Sigma_{\mathbf{z}}}
\nonumber \\
&=\frac{1}{2}\log \frac{ \det \left( \Sigma_{\mathbf{z}}+\Sigma_{\mathbf{x}}\right)}{ \det  \Sigma_{\mathbf{z}}}
=\frac{1}{2}\log \frac{ \det\left( \Sigma_{\mathbf{z}}+U_{\mathbf{x}}\Lambda_{\mathbf{x}}U^{\mathrm{T}}_{\mathbf{x}}
	\right)}{ \det  \Sigma_{\mathbf{z}}} =\frac{1}{2}\log \frac{ \det\left( \overline{\Sigma}_{\mathbf{z}}+\Lambda_{\mathbf{x}}
	\right)}{\det \overline{\Sigma}_{ \mathbf{z}}}, \nonumber
\end{flalign}
where $U_{\mathbf{x}} \Lambda_{\mathbf{x}} U^{\mathrm{T}}_{\mathbf{x}} $ is the eigen-decomposition of $\Sigma_{\mathbf{x}}$ with \begin{flalign}\Lambda_{\mathbf{x}} = \mathrm{diag} \left( \lambda_{1}, \ldots, \lambda_{m} \right), \nonumber
\end{flalign}
while  $\overline{\Sigma}_{\mathbf{z} }=U^{\mathrm{T}}_{\mathbf{x}}\Sigma_{\mathbf{z}} U_{\mathbf{x}} $. (Note that for a diagonal $\Sigma_{\mathbf{x}}$, we have $\lambda_{i} = \sigma_{\mathbf{x} \left( i \right)}^2$, where $\mathbf{x} \left( i \right)$ denotes the $i$-th element of $\mathbf{x}$, and $\sigma_{\mathbf{x} \left( i \right)}^2$ denotes its variance.) Note also that
\begin{flalign}
\mathrm{tr} \left( \overline{\Sigma}_{\mathbf{z}} \right)
=\mathrm{tr} \left(U^{\mathrm{T}}_{\mathbf{x}}\Sigma_{\mathbf{z} } U_{\mathbf{x}} \right)
= \mathrm{tr} \left(U_{\mathbf{x}} U^{\mathrm{T}}_{\mathbf{x}}\Sigma_{\mathbf{z} }  \right) =\mathrm{tr} \left(\Sigma_{\mathbf{z} }  \right)
= \mathbb{E} \left[ \sum_{i=1}^{m}\mathbf{z}^{2} \left( i \right) \right] \leq N.
\nonumber
\end{flalign}
It is known (see Lemma~3.2 of \cite{fang2017towards}) that
\begin{flalign}
\frac{1}{2} \log \frac{ \det\left(\overline{\Sigma}_{\mathbf{z} }+\Lambda_{\mathbf{x} } \right)}{\det \overline{\Sigma}_{\mathbf{z} } }
\geq \frac{1}{2} \log  \prod_{i=1}^m \left[\frac{ \overline{\sigma}_{\mathbf{z} \left( i \right)}^2+\lambda_{i}}{\overline{\sigma}_{\mathbf{z} \left( i \right)}^2} \right], \nonumber
\end{flalign}
where $\overline{\sigma}_{\mathbf{z} \left( i \right)}^2, i=1,\ldots,m$, are the diagonal terms of $\overline{\Sigma}_{\mathbf{z} }$, and the equality holds if $\overline{\Sigma}_{\mathbf{z} }$ is diagonal, whereas when $\overline{\Sigma}_{\mathbf{z} }$ is diagonal, we denote
\begin{flalign}
\overline{\Sigma}_{\mathbf{z} }=\mathrm{diag}\left(\overline{\sigma}_{\mathbf{z} \left(1\right)}^2,\ldots, \overline{\sigma}_{\mathbf{z} \left(m\right)}^2 \right)
=\mathrm{diag}\left(N_{1},\ldots,N_{m}\right) \nonumber
\end{flalign}
for simplicity.
Then, the problem reduces to that of choosing $N_1,\ldots, N_m$ to minimize 
\begin{flalign}
\frac{1}{2} \log  \prod_{i=1}^m \left( \frac{ N_{i}+\lambda_{i}}{N_{i}} \right)
= \sum_{i=1}^{m} \frac{1}{2}\log \left( 1+\frac{\lambda_{i}}{N_{i}} \right) \nonumber
\end{flalign}
subject to the constraint that 
\begin{flalign}
\sum_{i=1}^{m} N_{i} = \mathrm{tr} \left( \overline{\Sigma}_{\mathbf{z}} \right)  = N. \nonumber
\end{flalign}
Define the Lagrange function by
\begin{flalign}
\sum_{i=1}^{m} \frac{1}{2}\log \left( 1+\frac{\lambda_{i}}{N_{i}} \right)+\eta \left(\sum_{i=1}^{m} N_{i}-N\right), \nonumber
\end{flalign}
and differentiate it with respect to $N_{i}$, then we have
\begin{flalign}
\frac{\log \mathrm{e}}{2}\left( \frac{1}{N_{i}+ \lambda_{i}}-\frac{1}{N_{i}}\right) +\eta=0, \nonumber
\end{flalign}
or equivalently,
\begin{flalign}
N_{i}=\frac{\sqrt{\lambda_{i}^2+\zeta \lambda_{i}}-\lambda_{i}}{2}=\frac{\zeta}{2\left( \sqrt{1+\frac{\zeta }{\lambda_{i}}}+1
	\right)}, \nonumber
\end{flalign}
where $\zeta= 2 \log \mathrm{e}  / \eta \geq 0$ satisfies 
\begin{flalign}
\sum_{i=1}^{m} N_{i} = \sum_{i=1}^{m} \frac{\zeta}{2\left(\sqrt{1+\frac{\zeta }{\lambda_{i}}}+1
	\right)} = N. \nonumber
\end{flalign}
Consider now a scalar (dynamic) channel
\begin{flalign}
\mathbf{y}_{k} = \mathbf{x}_{k} + \mathbf{z}_{k}, \nonumber
\end{flalign}
where $ \mathbf{x}_{k},\mathbf{y}_{k},\mathbf{z}_{k} \in  \mathbb{R}$, and $ \left\{ \mathbf{z}_{k} \right\}$ is independent of $ \left\{ \mathbf{x}_{k} \right\} $. In addition, $\left\{ \mathbf{x}_{k} \right\}$ is stationary colored Gaussian with power spectrum $S_{\mathbf{x}} \left( \omega \right)$, and the noise power constraint is given by $\mathbb{E} \left[
\mathbf{z}^{2}_{k} \right] \leq N$.
%
We may then consider a block of consecutive uses from time $0$ to $k$ of this channel 
as $k+1$ channels in parallel \cite{Cov:06}. Particularly, let the eigen-decomposition of $\Sigma_{\mathbf{x}_{0,\ldots,k}}$ be given by
\begin{flalign}
\Sigma_{\mathbf{x}_{0,\ldots,k}}=U_{\mathbf{x}_{0,\ldots,k}}\Lambda_{\mathbf{x}_{0,\ldots,k}}U^{\mathrm{T}}_{\mathbf{x}_{0,\ldots,k}}, \nonumber
\end{flalign} 
where
\begin{flalign}
\Lambda_{\mathbf{x}_{0,\ldots,k}}
=\mathrm{diag} \left( \lambda_{0},\ldots,\lambda_{k} \right). \nonumber
\end{flalign}
Then, we have
\begin{flalign}
\min_{p_{\mathbf{z}_{0,\ldots,k}}:~\sum_{i=0}^{k} \mathbb{E}
	\left[ \mathbf{z}_{i}^{2} \right]
	\leq \left(k+1\right)N} \frac{I \left(\mathbf{x}_{0,\ldots,k}; \mathbf{y}_{0,\ldots,k} \right)}{k+1}
 =\frac{1}{k+1} \sum_{i=0}^{k} \frac{1}{2}\log \left( 1+\frac{\lambda_{i}}{N_{i} }\right), \nonumber
\end{flalign}
where
\begin{flalign}
N_{i}   = \frac{\zeta}{2\left(\sqrt{1+\frac{\zeta }{\lambda_{i}   }}+1
	\right)},~i=0,\ldots,k. \nonumber
\end{flalign}
Herein, $\zeta \geq 0$ satisfies
\begin{flalign}
\sum_{i=0}^{k} N_{i}  
= \sum_{i=0}^{k} \frac{\zeta}{2\left( \sqrt{1+\frac{\zeta }{\lambda_{i}  }}+1
	\right)}
= \left( k+1 \right) N, \nonumber
\end{flalign}
or equivalently,
\begin{flalign}
\frac{1}{k+1} \sum_{i=0}^{k} N_{i}  
= \frac{1}{k+1} \sum_{i=0}^{k} \frac{\zeta}{2\left( \sqrt{1+\frac{\zeta }{\lambda_{i}  }}+1
	\right)}
= N. \nonumber
\end{flalign}
In addition, as $k \to \infty$, the processes $\left\{ \mathbf{x}_{k} \right\}$, $\left\{ \mathbf{z}_{k} \right\}$, and $\left\{ \mathbf{y}_{k} \right\}$ are stationary, and
\begin{flalign}
\lim_{k\to \infty} \min_{p_{\mathbf{z}_{0,\ldots,k}}:~\sum_{i=0}^{k}
	\mathbb{E}
	\left[ \mathbf{z}_{i}^{2} \right] \leq \left(k+1\right)N} \frac{I \left(\mathbf{x}_{0,\ldots,k}; \mathbf{y}_{0,\ldots,k} \right)}{k+1} 
& =\inf_{p_{\mathbf{z}}}  \lim_{k\to \infty}\frac{I \left(\mathbf{x}_{0,\ldots,k}; \mathbf{y}_{0,\ldots,k} \right)}{k+1}\nonumber \\
&=\inf_{p_{\mathbf{z}}}  \limsup_{k\to \infty}\frac{I \left(\mathbf{x}_{0,\ldots,k}; \mathbf{y}_{0,\ldots,k} \right)}{k+1}
= \inf_{p_{\mathbf{z}}}I_{\infty} \left(\mathbf{x}; \mathbf{y} \right) \nonumber \\
&= L
. \nonumber
\end{flalign}
On the other hand, since the processes are stationary, the covariance
matrices are Toeplitz \cite{grenander1958toeplitz}, and their eigenvalues approach their limits as $k
\to \infty$. Moreover, the densities of eigenvalues on the real line
tend to the power spectra of the processes \cite{gutierrez2008asymptotically}. Accordingly,
\begin{flalign}
L
&=\lim_{k\to \infty} \min_{p_{\mathbf{z}_{0,\ldots,k}}:~\sum_{i=0}^{k}
	\mathbb{E}
	\left[ \mathbf{z}_{i}^{2} \right] \leq \left(k+1\right) N} \frac{I \left(\mathbf{x}_{0,\ldots,k}; \mathbf{y}_{0,\ldots,k} \right)}{k+1} 
= \lim_{k\to \infty} \frac{1}{k+1} \sum_{i=0}^{k}\frac{1}{2}  \log \left( 1+\frac{\lambda_{i}}{N_{i}} \right)
\nonumber \\
&= \frac{1}{2\pi} \int_{-\pi}^{\pi} \frac{1}{2}  \log \left[ 1+\frac{S_{\mathbf{x}} \left(\omega \right)}{N \left(\omega \right)} \right] \mathrm{d} \omega 
= \frac{1}{2\pi} \int_{-\pi}^{\pi} \log \sqrt{ 1+\frac{S_{\mathbf{x}} \left(\omega \right)}{N \left(\omega \right)} } \mathrm{d} \omega,
\nonumber
\end{flalign}
where
\begin{flalign}
N \left(\omega \right) 
= \frac{\zeta}{2\left[\sqrt{1+\frac{\zeta }{S_{\mathbf{x}} \left( \omega \right)}}+1 \right]}, \nonumber
\end{flalign}
and $\zeta \geq 0 $ satisfies
\begin{flalign}
\lim_{k\to \infty} \frac{1}{k+1}\sum_{i=0}^{k} N_{i}
=\frac{1}{2\pi} \int_{-\pi}^{\pi} N \left(\omega \right) \mathrm{d}  \omega \nonumber 
=N. \nonumber
\end{flalign}
This concludes the proof.

\section{Proof of Theorem~\ref{fading}}
In general, we can consider a block of consecutive uses of this channel (from time $0$ to $k$)
as $k+1$ channels in parallel \cite{Cov:06}.

Note first that without noise side information, the variance $\sigma_{\mathbf{x}_{i}}^2 = \left| h_{i} \right|^2 \sigma_{\widehat{\mathbf{x}}}^2 $ of $\mathbf{x}_{i}$ at time $i, i = 0, \ldots, k$, is unknown to the noise designer; note that herein $ h_{0, \ldots, k}$ denote the realizations of $ \mathbf{h}_{0, \ldots, k}$. Accordingly, the best the noise designer can do is to make the noise $\left\{ \mathbf{z}_k \right\}$ stationary white Gaussian that is independent of $\left\{ \widehat{\mathbf{x}}_k\right\}$ and $\left\{ \mathbf{h}_k\right\}$ and with variance $N$ (cf. discussions on fast fading channel capacity without transmitter side information \cite{goldsmith2005wireless, Tse:04}). 
Particularly, it holds that
\begin{flalign}
\min_{p_{\mathbf{z}_{0,\ldots,k}}:~\sum_{i=0}^{k} \mathbb{E}
	\left[ \mathbf{z}_{i}^{2} \right]\leq \left(k+1\right)N} \frac{I \left(\mathbf{x}_{0,\ldots,k}; \mathbf{y}_{0,\ldots,k} \right)}{k+1} =\frac{1}{k+1} \sum_{i=0}^{k} \frac{1}{2} \log \left(1+\frac{\left| h_i \right|^2 \sigma_{\mathbf{x}}^2 }{N}\right).\nonumber
\end{flalign}
Correspondingly,
\begin{flalign}
\lim_{k\to \infty} \min_{p_{\mathbf{z}_{0,\ldots,k}}:~\sum_{i=0}^{k} \mathbb{E}
	\left[ \mathbf{z}_{i}^{2} \right]\leq \left(k+1\right)N} \frac{I \left(\mathbf{x}_{0,\ldots,k}; \mathbf{y}_{0,\ldots,k} \right)}{k+1}
&= \lim_{k\to \infty} \frac{1}{k+1} \sum_{i=0}^{k} \frac{1}{2} \log \left(1+\frac{\left| h_i \right|^2 \sigma_{\mathbf{x}}^2 }{N}\right) \nonumber \\
& = \mathbb{E} \left[ \frac{1}{2} \log \left(1+\frac{\left| \mathbf{h}_i \right|^2 \sigma_{\mathbf{x}}^2 }{N}\right) \right]
.\nonumber
\end{flalign}
On the other hand, with the aforementioned optimal $\left\{ \mathbf{z}_k \right\}$,
\begin{flalign}
\lim_{k\to \infty} \min_{p_{\mathbf{z}_{0,\ldots,k}}:~\sum_{i=0}^{k}
	\mathbb{E}
	\left[ \mathbf{z}_{i}^{2} \right] \leq \left(k+1\right)N} \frac{I \left(\mathbf{x}_{0,\ldots,k}; \mathbf{y}_{0,\ldots,k} \right)}{k+1}
& =\inf_{p_{\mathbf{z}}}  \lim_{k\to \infty}\frac{I \left(\mathbf{x}_{0,\ldots,k}; \mathbf{y}_{0,\ldots,k} \right)}{k+1}\nonumber \\
& =\inf_{p_{\mathbf{z}}}  \limsup_{k\to \infty}\frac{I \left(\mathbf{x}_{0,\ldots,k}; \mathbf{y}_{0,\ldots,k} \right)}{k+1}
= \inf_{p_{\mathbf{z}}}I_{\infty} \left(\mathbf{x}; \mathbf{y} \right) \nonumber \\
& = L
. \nonumber
\end{flalign}
As such,  
\begin{flalign}
L = \mathbb{E} \left[ \frac{1}{2} \log \left(1+\frac{\left| \mathbf{h}_k \right|^2 \sigma_{\mathbf{x}}^2 }{N}\right) \right]. \nonumber
\end{flalign}

Note next that with noise side information, the variance $\sigma_{\mathbf{x}_{i}}^2 = \left| h_{i} \right|^2 \sigma_{\widehat{\mathbf{x}}}^2 $ of $\mathbf{x}_{i} = \mathbf{h}_{i} \widehat{\mathbf{x}}_{i}$ at time $i$ is known to the noise designer. Accordingly, the noise designer can make use of this additional information (compared to the previous case without noise side information) and the optimal noise $\left\{ \mathbf{z}_k \right\}$ is white Gaussian that is independent of $\left\{ \widehat{\mathbf{x}}_k\right\}$ and $\left\{ \mathbf{h}_k\right\}$ and with a time-varying variance $N_{i}$ as a function of the realization $h_{i}$ of $\mathbf{h}_{i}$ (cf. discussions on fast fading channel capacity with transmitter side information \cite{goldsmith2005wireless, Tse:04}). 
Particularly, we can consider a block of consecutive uses of this channel (from time $0$ to $k$)
as $k+1$ channels in parallel \cite{Cov:06}, and it holds  that
\begin{flalign}
\min_{p_{\mathbf{z}_{0,\ldots,k}}:~\sum_{i=0}^{k} \mathbb{E}
	\left[ \mathbf{z}_{i}^{2} \right]\leq \left(k+1\right)N} \frac{I \left(\mathbf{x}_{0,\ldots,k}; \mathbf{y}_{0,\ldots,k} \right)}{k+1} =\frac{1}{k+1} \sum_{i=0}^{k}  \left(1+\frac{\left| h_i \right|^2 \sigma_{\mathbf{x}}^2 }{N_{i}}\right),\nonumber
\end{flalign}
where
\begin{flalign}
N_{i} = \frac{\zeta}{2\left(\sqrt{1+\frac{\zeta }{\left| h_i \right|^2 \sigma_{\mathbf{x}}^2}}+1
	\right)}. \nonumber
\end{flalign}
Herein, $\zeta \geq 0$ satisfies
\begin{flalign}
\sum_{i=0}^{k} N_{i}
= \sum_{i=0}^{k} \frac{\zeta}{2\left(\sqrt{1+\frac{\zeta }{\left| h_i \right|^2 \sigma_{\mathbf{x}}^2}}+1
	\right)}
= \left( k + 1 \right)N,\nonumber
\end{flalign}
or equivalently,
\begin{flalign}
\frac{1}{k+1} \sum_{i=0}^{k} N_{i}
= \frac{1}{k+1} \sum_{i=0}^{k} \frac{\zeta}{2\left(\sqrt{1+\frac{\zeta }{\left| h_i \right|^2 \sigma_{\mathbf{x}}^2}}+1
	\right)}
=N.\nonumber
\end{flalign}
Hence, as $k \to \infty$, it holds that
\begin{flalign}
\lim_{k\to \infty} \min_{p_{\mathbf{z}_{0,\ldots,k}}:~\sum_{i=0}^{k} \mathbb{E}
	\left[ \mathbf{z}_{i}^{2} \right]\leq \left(k+1\right)N} \frac{I \left(\mathbf{x}_{0,\ldots,k}; \mathbf{y}_{0,\ldots,k} \right)}{k+1} 
& =\lim_{k\to \infty}\frac{1}{k+1} \sum_{i=0}^{k}  \left(1+\frac{\left| h_i \right|^2 \sigma_{\mathbf{x}}^2 }{N_{i}}\right) \nonumber \\
& = \mathbb{E} \left[ \frac{1}{2}\log
\left(1+\frac{ \left| \mathbf{h}_k \right|^2 \sigma_{\mathbf{x}}^2}{N_{k}}\right) \right],\nonumber
\end{flalign}
where
\begin{flalign}
N_{k} = \frac{\zeta}{2\left(\sqrt{1+\frac{\zeta }{\left| h_k \right|^2 \sigma_{\mathbf{x}}^2}}+1
	\right)}, \nonumber
\end{flalign}
and $\zeta \geq 0$ satisfies
\begin{flalign}
\lim_{k\to \infty} \frac{1}{k+1} \sum_{i=0}^{k} N_{i}
= \mathbb{E} \left[   N_{k} \right]=N.\nonumber
\end{flalign}
On the other hand,  with the aforementioned optimal $\left\{ \mathbf{z}_k \right\}$,
\begin{flalign}
\lim_{k\to \infty} \min_{p_{\mathbf{z}_{0,\ldots,k}}:~\sum_{i=0}^{k}
	\mathbb{E}
	\left[ \mathbf{z}_{i}^{2} \right] \leq \left(k+1\right)N} \frac{I \left(\mathbf{x}_{0,\ldots,k}; \mathbf{y}_{0,\ldots,k} \right)}{k+1} 
& =\inf_{p_{\mathbf{z}}}  \lim_{k\to \infty}\frac{I \left(\mathbf{x}_{0,\ldots,k}; \mathbf{y}_{0,\ldots,k} \right)}{k+1}\nonumber \\
& =\inf_{p_{\mathbf{z}}}  \limsup_{k\to \infty}\frac{I \left(\mathbf{x}_{0,\ldots,k}; \mathbf{y}_{0,\ldots,k} \right)}{k+1}
= \inf_{p_{\mathbf{z}}}I_{\infty} \left(\mathbf{x}; \mathbf{y} \right) \nonumber \\
&= L
. \nonumber
\end{flalign}
As such,  
\begin{flalign}
L = \mathbb{E} \left[ \frac{1}{2} \log \left(1+\frac{\left| \mathbf{h}_k \right|^2 \sigma_{\mathbf{x}}^2 }{N_k}\right) \right]. \nonumber
\end{flalign}
This concludes the proof.

%
%
%
%
%
%
%
%
%
%
%
%
%
%
%


\bibliographystyle{IEEEtran}
\bibliography{references}
%
%
%
%
%
%
\end{document}